\documentclass[longauth]{aa}
\usepackage{graphicx}
\usepackage{txfonts}
\usepackage{amsmath}	
\usepackage{amssymb}	
\usepackage{mathtools}
\usepackage{threeparttable}
\usepackage{longtable}
\usepackage{tabto}
\usepackage{adjustbox}
\usepackage{multicol} 
\usepackage{multirow} 
\usepackage[T1]{fontenc}
\usepackage[utf8]{inputenc}

\usepackage{color}
\usepackage{natbib,twoopt}
\usepackage[hyphenbreaks]{breakurl}
\usepackage[breaklinks]{hyperref}      
\bibpunct{(}{)}{;}{a}{}{,}             
\definecolor{cobalt}{rgb}{0.06, 0.2, 0.65}
\hypersetup{
  colorlinks,
  citecolor=blue,
  linkcolor=blue,
  urlcolor=blue,
}
\makeatletter
  \newcommandtwoopt{\citeads}[3][][]{\href{http://adsabs.harvard.edu/abs/#3}%
    {\def\hyper@linkstart##1##2{}%
     \let\hyper@linkend\@empty\citealp[#1][#2]{#3}}}
  \newcommandtwoopt{\citepads}[3][][]{\href{http://adsabs.harvard.edu/abs/#3}%
    {\def\hyper@linkstart##1##2{}%
     \let\hyper@linkend\@empty\citep[#1][#2]{#3}}}
  \newcommandtwoopt{\citetads}[3][][]{\href{http://adsabs.harvard.edu/abs/#3}%
    {\def\hyper@linkstart##1##2{}%
     \let\hyper@linkend\@empty\citet[#1][#2]{#3}}}
  \newcommandtwoopt{\citeyearads}[3][][]%
    {\href{http://adsabs.harvard.edu/abs/#3}
    {\def\hyper@linkstart##1##2{}%
     \let\hyper@linkend\@empty\citeyear[#1][#2]{#3}}}
\makeatother

\newcommand{\Msun}{M$_{\odot}$}
\newcommand{\Rsun}{R$_{\odot}$}
\newcommand{\Msunyr}{M$_{\odot}$\,yr$^{-1}$}

\definecolor{smalt(darkpowderblue)}{rgb}{0.0, 0.2, 0.6}
\definecolor{forestgreen(traditional)}{rgb}{0.0, 0.5, 0.0}

\defcitealias{RVJ}{RVJ}

\defcitealias{MD17}{MD17}

\newcommand{\gaia}{\textit{Gaia}}

\newcommand{\paradoxical}{SDSS\,J1257$+$5428}

\newcommand{\VAnd}{V479\,And}
\newcommand{\VSgr}{V1082\,Sgr}

\usepackage[dvipsnames]{xcolor}

\begin{document} 

   \title{Revisiting the extremely long-period cataclysmic variables \\ V479 Andromedae and V1082 Sagitarii}

\titlerunning{Revisiting the extremely long-period CVs V479 And and V1082 Sgr}


    \author{
    Gagik Tovmassian\inst{1}\thanks{\email{gag@astro.unam.mx}}
    \and
    Diogo Belloni\inst{2}
    \and
    Anna F. Pala\inst{3}
    \and
    Thomas Kupfer\inst{4,5}
    \and
    Weitian Yu\inst{4}
    \and
    Boris~T.~G\"ansicke\inst{6}
    \and
    Elizabeth O. Waagen\inst{7}
    \and
    Juan-Luis González-Carballo\inst{8}
    \and
    Paula Szkody\inst{9}
    \and
    Domitilla de Martino\inst{10}
    \and
    Matthias R. Schreiber\inst{11}
    \and
    Knox S. Long\inst{12,13}
    \and
    Alan Bedard\inst{7}
    \and
    Slawomir Bednarz\inst{14}
    \and
    Jordi Berenguer\inst{15}
    \and
    Krzysztof Bernacki\inst{14}
    \and
    Simone Bolzoni\inst{16}
    \and
    Carlos Botana-Albá\inst{17}
    \and
    Christopher Cantrell\inst{7}
    \and
    Walt Cooney\inst{18}
    \and
    Charles Cynamon\inst{19}
    \and
    Pablo De la Fuente Fernández\inst{20}
    \and
    Sjoerd Dufoer\inst{21}
    \and
    Esteban Fernández Mañanes\inst{22}
    \and
    Faustino García-Cuesta\inst{23}
    \and
    Rafael Gonzalez Farfán\inst{24}
    \and
    Pierre A. Fleurant\inst{25}
    \and
    Enrique A. Gómez\inst{26}
    \and
    Matthew J. Green\inst{27}
    \and
    Franz-Josef Hambsch\inst{28,29,30,7}
    \and
    Penko Jordanov\inst{31}
    \and
    Emmanuel Kardasis\inst{32}
    \and
    David Lane\inst{16}
    \and
    Darrell Lee\inst{7} 
    \and
    Isabel J. Lima\inst{33,34}
    \and
    Fernando Limón Martínez\inst{35}
    \and
    Gianpiero Locatelli\inst{36}
    \and
    Jose-Luis Martin-Velasco\inst{37}
    \and
    Daniel J. Mendicini\inst{38}
    \and
    Michel Michaud\inst{39}
    \and
    Moisés Montero Reyes Ortíz\inst{40}
    \and
    Mario Morales Aimar\inst{41}
    \and
    Gordon Myers\inst{42}
    \and
    Ramon Naves Nogues\inst{43}
    \and
    Giuseppe Pappa\inst{44}
    \and
    Andrew Pearce\inst{45}
    \and
    James Pierce\inst{46}
    \and
    Adam Popowicz\inst{14}
    \and
    Claudia V. Rodrigues\inst{47}
    \and
    Nieves C. Rodríguez\inst{48}
    \and
    David Quiles Amat\inst{7}
    \and
    Esteban Reina-Lorenz\inst{49}
    \and
    José-Luis Salto-González\inst{50}
    \and
    Jeremy Shears\inst{51}
    \and
    John Sikora\inst{52,53}
    \and
    André Steenkamp\inst{54}
    \and
    Rod Stubbings\inst{55}
    \and
    Brad Young\inst{56}
    \and
    Ivan L. Walton\inst{51}}

  \authorrunning{Tovmassian, et al.}

    \institute{
    Universidad Nacional Aut\'onoma de M\'exico, Instituto de Astronom\'{i}a, Aptdo Postal 106, Ensenada 22860, Baja California, M\'exico
    \and
    S\~ao Paulo State University (UNESP), School of Engineering and Sciences, Guaratinguet\'a, Brazil
    \and
    European Southern Observatory, Karl Schwarzschild Stra{\ss}e 2, D-85748 Garching, Germany
    \and
    Hamburger Sternwarte, University of Hamburg, Gojenbergsweg 112, 21029 Hamburg, Germany
    \and
    Department of Physics and Astronomy, Texas Tech University, 2500 Broadway, Lubbock, TX 79409, USA
    \and
    Department of Physics, University of Warwick, Coventry, CV4 7AL, UK
    \and
    Association of Variable Star Observers (AAVSO), 185 Alewife Brook Pkwy, Suite 410, Cambridge, MA 02138, USA
    \and
    Observadores de Supernovas (ObSN), Observatorio Cerro del Viento, MPC I84, Pl. Fernández Pirfano 3-5A, Badajoz, 06010, Spain
    \and
    Department of Astronomy, University of Washington, Seattle, WA 98195 USA
    \and
    INAF - Osservatorio Astronomico di Capodimonte, Salita Moiariello 16, 80131, Naples, Italy
    \and
    Departamento de F{\'i}sica, Universidad T{\'e}cnica Federico Santa María, Av. Espa{\~n}a 1680, Valpara{\'i}so, Chile
    \and
    Space Telescope Science Institute, 3700 San Martin Drive, Baltimore, MD, 21218, USA
    \and
    Eureka Scientific, Inc., 2452 Delmer Street Suite 100, Oakland, CA 94602-3017, USA
    \and
    Silesian University of Technology, Akademicka 16, Gliwice, Poland
    \and
    Observadores de Supernovas (ObSN), Observatorio Major de Dalt, C/ Major 54-2, Sant Celoni, Barcelona, 08470, Spain
    \and
    Abbey Ridge Observatory, 45 Abbey Rd, Stillwater Lake, NS, B3Z1R1 Canada
    \and
    Observadores de Supernovas (ObSN), Observatorio Magalofes, MPC Y85, C/ Feal 20, Fene, A Coruña, 15509, Spain
    \and
    Madrona Peak Observatory, 4635 Shadow Grass Dr, Katy, TX 77493, USA
    \and
    Supra Solem Observing, SkiesAway Remote Observatory, Bradley, CA
    \and
    Observadores de Supernovas (ObSN), Observatorio El Llagarín, C/ Rio Dobra 18, Oviedo 33010, Spain
    \and
    Vereniging voor Sterrenkunde, Zeeweg 96, 8200 Brugge, Belgium
    \and
    Observadores de Supernovas (ObSN), Observatorio Estelia, MPC Y90, C/ Ladines 12, Ladines, Asturias, 33993, Spain
    \and
    Observadores de Supernovas (ObSN), Observatorio LaVara, MPC J38, Barrio La Vara s/n, Valdés, Asturias, 33784, Spain
    \and
    Observadores de Supernovas (ObSN), Observatorio Uraniborg  C/ Antequera, 8, 41400 Écija, Sevilla, Spain
    \and
    Mittelman ATMoB Observatory/Amateur Telescope Makers of Boston, Inc. 99 College Ave. Arlington, Massachusetts 02474 USA
    \and
    Dept. Chemistry and Physics, College of Arts and Sciences, Western Carolina University, Apodaca 116B,1 University Way, Cullowhee, NC 28723
    \and
    Homer L. Dodge Department of Physics and Astronomy, University of Oklahoma, 440 W. Brooks Street, Norman, OK 73019, USA
    \and
    Vereniging Voor Sterrenkunde (VVS), Zeeweg 96, 8200 Brugge, Belgium 
    \and
    Groupe Europ\'een d’Observations Stellaires (GEOS), 23 Parc de Levesville, 28300 Bailleau l\'\,Eveque, France
    \and 
    Bundesdeutsche Arbeitsgemeinschaft f\"ur Ver\"anderliche Sterne (BAV), Munsterdamm 90, 12169 Berlin, Germany
    \and
    American Association of Variable Star Observers (AAVSO), bul. Rakovski 39-V-14, 6400, Dimitrovgrad, Bulgaria
    \and
    Pelagia-Eleni observatory, Glyfada, Athens, Greece
    \and
    CONICET-Universidad de Buenos Aires, Instituto de Astronom\'{i}a y F\'{i}sica del Espacio (IAFE), Av. Inte. G\"{u}iraldes 2620, C1428ZAA, Buenos Aires, Argentina 
    \and
    Universidade Estadual Paulista “Júlio de Mesquita Filho”, UNESP, Campus of Guaratinguetá, Av. Dr. Ariberto Pereira da Cunha, 3334 - Pedregulho, Guaratinguetá - SP, 12516-410, Brazil5
    \and
    Observadores de Supernovas (ObSN), Observatorio Mazariegos, MPC Z50, Mazariegos, Palencia, 34170, Spain
    \and
    Observadores de Supernovas (ObSN), Maritime Alps Observatory, MPC K32, Via Mellana 26, Cuneo, 12100, Italy
    \and
    Observadores de Supernovas (ObSN), Observatorio Carpe Noctem, MPC I72, Paseo de la maliciosa 11, Collado Mediano, Madrid, 28027, Spain
    \and
    Sección de Estrellas Variables - Centro de Observadores del Espacio – Liga Iberoamericana de Astronomía. (S.E.V.-C.O.D.E./L.I.A.D.A.), Av. Almirante Guillermo Brown Nro. 4998, Costanera Oeste 3000, Santa Fe, Argentina
    \and
    MCD Observatory, 23 Langlois, G0K1H0, Canada
    \and
    Fundación Astronomía Sigma Octante, Pasaje Man Cesped 592, 0301, Bolivia
    \and
    Observadores de Supernovas (ObSN), Observatorio de Sencelles, MPC K14, Camí de Sonfred 1, Sencelles, Islas Baleares, 07140, Spain
    \and
    American Association of Variable Star Observers (AAVSO), 5 Inverness Way, Hillsborough, CA 94010 USA
    \and
    Observadores de Supernovas (ObSN), Observatorio Montcabrer, MPC 213, C/ Jaume Balmes 24, Cabrils, Barcelona, 08348, Spain
    \and
    American Association of Variable Star Observers (AAVSO), Via Gioacchino Rossini 2b, 95030, Pedara, Italy
    \and
    Nedlands Observatory, 35 Viewway, Nedlands WA 6009, Australia
    \and
    Toadhall Observatory, 2095, Bakery Hill, Victoria, 3354, Australia
    \and
    Instituto Nacional de Pesquisas Espaciais (INPE/MCTI), Av. dos Astronautas, 1758, S\~ao Jos\'e dos Campos, SP, Brazil
    \and
    Instituto de Astrofísica de Canarias, C/O Vía Láctea s/n, 38200 La Laguna, Spain
    \and
    Observadores de Supernovas (ObSN), Observatorio de Masquefa, MPC 232, Av. Can Marcet 41, Masquefa, Barcelona, 08783, Spain
    \and
    Observadores de Supernovas (ObSN), Cal Maciarol mòdul 8 Observatory, MPC A02, Masia Cal Maciarol, Camí de l'Observatori s/n, Àger, Lleida, 25691, Spain
    \and
    British Astronomical Association Variable Star Section, P.O. Box 702, Tonbridge, TN9 9TX, United Kingdom
    \and
    Dark Sky New Mexico, 30 Washburn Rd., Animas, NM, 88020, USA
    \and
    Lake County Astronomical Association, 28478 W. Brandenburg Road, Ingleside, IL 60041, USA
    \and
    Southwater Observatory, Horsham, West Sussex, UK
    \and
    American Association of Variable Stars Observers (AAVSO), Tetoora Road Observatory, 2643 Warragul-Korumburra Road, Tetoora Road, 3821, Australia
    \and
    International Astronomical Union Centre for the Protection of the Dark and Quiet Sky from Satellite Constellation Interference, NOIRLab , 950 N. Cherry Ave, Tucson, AZ 85719, USA}

   \date{Received ; accepted }

 
  \abstract
   {The overwhelming majority of cataclysmic variables (CVs) have orbital periods shorter than 10 hr. However, a few have much longer orbital periods, and their formation and existence pose certain challenges for the CV evolution models. These extremely long-period CVs must host nuclearly evolved donor stars (i.e., subgiants), as otherwise, the companion of the white dwarf would be too small to fill its Roche lobe. This makes them natural laboratories for testing binary evolution models and accretion processes with subgiant donors, with applications extending beyond white dwarf binaries. Despite the importance of compact objects accreting from subgiant donors, we still do not fully understand how they form and evolve.}
    {To shed light on the formation and evolution of accreting compact objects with subgiant companions, we investigated two extremely long-period CVs in detail, namely {V479\,And} ($P_\mathrm{orb} \simeq 14\,$hr) and {V1082\,Sgr} ($P_\mathrm{orb} \simeq 21\,$hr). We searched for reasonable formation pathways to explain their refined stellar and binary parameters.}
   {We used a broad set of new observations, including ultraviolet and infrared spectroscopy, results of circular polarimetry, and improved \gaia\ DR3 distance estimates to determine fundamental parameters (e.g. effective temperatures, masses, and radii of the donor stars) to be confronted with numerical simulations. Furthermore, we utilized the MESA code to conduct numerical simulations, employing state-of-the-art prescriptions, such as the Convection And Rotation Boosted (CARB) model for strong magnetic braking. }
   {Our observations reaffirm that V479\,And is a polar and show that V1082\,Sgr is an intermediate polar.  Both systems have unusual chemical composition and very low masses for their assigned spectral classes. This most likely indicates that they underwent thermal timescale mass transfer. We found models for both extremely long-period CVs that can reasonably reproduce their properties. CV evolution needs to be convergent (i.e., toward shorter orbital periods), which is only possible if the magnetic braking is sufficiently strong. }
   {We conclude that the donor stars in both V479\,And and V1082\,Sgr are filling their Roche lobes, ruling out previous models in which they are under-filling their Roche lobes. Our findings suggest that orbital angular momentum loss is stronger due to magnetic braking in CVs with subgiant donors compared to those with unevolved donors. In addition, our findings suggest that extremely long-period CVs could significantly contribute to the population of double white dwarf binaries in close orbits (orbital periods ${\lesssim1}$~d).}


   \maketitle
%
\section{Introduction}
Cataclysmic variables (CVs) are interacting binary stars in which a white dwarf (WD) accretes material from a low-mass, near-main-sequence companion star.  The long-term evolution of CVs is driven by the secular angular momentum loss (AML), which provides conditions to maintain them in a semi-detached configuration. 
In the standard model of CV evolution, the dominant AML mechanism in long-period systems (${P_{\rm orb}\ge3}$~hr) is thought to be the magnetic braking (MB). Meanwhile, short-period CVs (${P_{\rm orb}\le3}$~hr) are assumed to be driven by AML caused by gravitational radiation emission. 
However, this model is incomplete since it cannot account for the main CV observables (i.e., orbital period distribution, orbital period minimum, space density, and WD mass distribution).
Most likely, other sources of AMLs, such as consequential AML and/or residual MB below the orbital period gap, are not taken into account, which are required to better explain these observables  \citep[e.g.,][]{Knigge_2011_OK,Schreiber_2016,Belloni_2018b,Pala_2022,BelloniSchreiberREVIEW}.

Struggles in developing fully consistent evolutionary models for CV evolution are likely related to the fact that observational samples are still strongly biased against fainter systems \citep{Schreiberetal24-1}.
Therefore, larger and less biased samples are required to properly compare simulations and observations.
The best way to progress further is by creating highly complete and large volume-limited samples. Large observational efforts towards providing such samples are currently ongoing, and the first results are promising 
\citep[][]{Pala_2020,Inight_2023,Inight2025}.

Most CVs host M dwarf donors and the small scatter in the radius-mass relation of this population provides strong evidence for a tight evolutionary pathway not affected by the evolutionary stage of the donor stars \citep{McAllister_2019}.
However, around ${5-15}$\,\%  \citep{Gaensicke2003,Pala_2020} of the CVs host nuclear evolved donors, with a much greater scatter in the radius-mass relation, which is most likely due to the different degrees of evolution of the donor stars.
Until recently, the evolution of CVs with nuclear evolved donors has been investigated with the so-called RVJ prescription for magnetic braking \citep*[][]{RVJ}. 
This prescription, though, does not work generally. It provides AML rates that are too low to explain observations not only of accreting WDs but also of accreting neutron stars \citep[e.g.,][]{Van_2019} and accreting black holes \citep[e.g.,][]{Wiktorowicz_2014}.
It also fails to explain the existence of close detached millisecond pulsars with extremely low-mass WDs \citep[e.g.,][]{Istrate_2014}, which are descendants of low-mass X-ray binaries.

A more promising magnetic braking prescription, the Convection And Rotation Boosted (CARB) model, was proposed by \citet{CARB}, who derived a self-consistent prescription that takes into account wind mass loss, rotation, and the generation of the magnetic field due to motions in the convective zone.
This model provides significantly stronger torques and was put forward to solve the disagreement between observed and predicted (by the RVJ prescription) mass transfer rates in neutron star low-mass X-ray binaries, the predictions by the RVJ model being at least an order of magnitude lower \citep[e.g.,][]{Deng_2021}.
The CARB model was later applied to progenitors of millisecond pulsar binaries with extremely low-mass WDs by \citet{Soethe_2021}, who showed that they could be formed without facing any fine-tuning problem.
More recently, \citet{BelloniSchreiber_2023} applied the CARB model to progenitors of AM\,CVn binaries and showed that the main problems faced by the CV channel (i.e. difficulty to form them and detectable amounts of hydrogen) are solved and that CVs with nuclear evolved donors might even be the dominant population among progenitors of AM\,CVn binaries. 

A population of CVs with nuclear evolved donors that have been overlooked are those with extremely long orbital periods.
The overwhelming majority of CVs, including those with nuclear evolved donors, have orbital periods ${\le10}$~hr \citep[][]{Gansicke_2009,2023MNRAS.524.4867I}.
Among the few systems with longer orbital periods, the evolutionary status of two particular CVs, which are V479\,Andromedae and V1082\,Sagitarii (hereafter \VAnd~and \VSgr) with orbital periods of 14.26 and 20.82\,hr, 
has been a source of debate for years
\citep{2002ASPC..261..102S,Tovmassian_2016,2019MNRAS.489.3031X,2020MNRAS.492.4344S}. 

Another potential issue regarding CVs with extremely long orbital periods is the large fraction of magnetic white dwarfs in known systems. V479\,And was proven to harbor a strongly magnetic white dwarf from the X-ray light curve \citep{2013A&A...553A..28G}. Recently, \citet{2025ApJ...984..152L} measured the circular polarization of V1082\,Sgr and demonstrated its presence with an amplitude of less than 1\%. The modulation in the polarization curve was interpreted as the spin period of a magnetic WD, produced by the cyclotron emission from the post-shock region near the WD surface. 
The estimated magnetic field strength of $\sim10^{7}$~G and temperatures that reach up to 17~keV at the top of the postshock region are consistent with the estimate using X-ray data by \citet{Bernardini_2013}. 

Here, we present new observations of these two CVs that largely improve previous observational characterizations and present evolutionary models that can naturally explain both systems as extremely long-period CVs with nuclear evolved donors.
We show that they host subgiant donors, underwent thermal timescale mass transfer, and that CV evolution has to be convergent to explain their properties, which occurs if the magnetic braking torques are stronger than those required to explain CVs with unevolved main-sequence stars.

\section{New observations}
\subsection{HST COS spectroscopy}

V479\,And was observed using the Cosmic Origin Spectrograph (COS) as part of a large \textit{HST} program in Cycle 30 (GO-16659, PIs: A.F. Pala and T. Kupfer) on 2022-12-22/23.
The data were collected for a total exposure time of 9317.76\,s through the Primary Science Aperture. The far-ultraviolet (FUV) channel, covering the wavelength range $800~\AA<\lambda<1950~\AA$, and the G140L grating, at the central wavelength of 800~\AA~and with a nominal resolving power of R $\simeq 3000$, were used. 
The data were collected in TIME–TAG mode, i.e., recording the time of arrival and the position on the detector of each detected photon, allowing the construction of ultraviolet light curves.
The \textit{HST}, SDSS spectra, as well as Vizier data for V479\,And were deredened according to \citet{1999PASP..111...63F} using {\sl unred} procedure included in  {\sl Python AstroLib} packages. A color excess E(B-V)=0.07 \citep{Green_2019, Schlegel_1998} with standard reddening ratio R$_{\mathrm{V}}=3.1$ were used. 

The COS FUV detector consists of a photon-counting microchannel plate that converts the incoming photons into electronic pulses. An excessive photon flux could result in permanent damage, and even the loss, of the detector. 
CVs are characterised by quiescent states of low-mass transfer rates onto the WD, interrupted by bright disc outbursts, during which the binary typically brightens by 2-8 magnitudes \citep{Maza+1983,Warner1995,Templeton2007}. These outbursts last for a few days or weeks. Their recurrence times vary widely among CVs and do not depend directly on the orbital period. During the outbursts, the disc is the dominant source of emission even in the far ultraviolet and becomes sufficiently bright to potentially damage the COS detector. In order to ensure that COS spectroscopy was obtained during quiescence and our targets were safe to be observed, we have obtained intensive continuous ground-based monitoring up until 24 h prior to the COS observations. 
We carried out this monitoring programme using the Las Cumbres Observatory (LCO), in close collaboration with the global citizen scientist community, including the American Association of Variable Star Observers (AAVSO) and the Observadores de Supernovas (ObSN). Only their outstanding support has made the \textit{HST} observations of V479\,And possible. In selected cases, we applied for an approximately 1000\,sec exposure with either UV filter of the UVOT telescope \citep{2005SSRv..120...95R} on board of \textit{Swift} spacecraft \citep{2004ApJ...611.1005G} to get a good idea of the UV flux of the objects programmed for observation and ensure safety of COS detectors. 

Unfortunately, in the case of V1082\,Sgr, the monitoring revealed that the object was in a high state at the time of the scheduled \textit{HST} observation. As the increased ultraviolet flux could have damaged the \textit{HST} COS detectors, the observations were aborted and no data were collected for this object.

\subsection{TESS photometry }

V479 And was observed with the \textit{Transiting Exoplanet Survey Satellite} (\textit{TESS}) \citep{Ricker_2015} on 2019-10-08 (Sector 17), 2022-09-30 (Sector 57), and 20-24-27 (Sector 84) under programs G05094, G022071, and G07025. For each sector, \textit{TESS} continuously collects photometric data for about 28 days, except for a gap of about 2–3 days in the middle, near perigee. Our data consists of three sectors, about $\sim1100$ and $\sim700$\ days apart.  The data were extracted from the TESS database 
through the Mikulski Archive for Space Telescopes (MAST) High-Level Science Products  (HLSP).  The \textit{TESS} bandpass is similar to a combined R and I filter. All of the photometry presented here was obtained at 120-s cadence. 

For the photometric analysis, we employed the \texttt{PDCSAP} fluxes (Pre-search Data Conditioning Simple Aperture Photometry) provided by the TESS pipeline. These fluxes are corrected for known instrumental systematics---such as spacecraft jitter, scattered light, and thermal effects---and thus provide more reliable light curves than the raw \texttt{SAP} fluxes. We converted these fluxes to TESS magnitudes using the relation recommended in the TESS Instrument Handbook \citep{Vanderspek2018}:
\(
m_{\mathrm{TESS}} = -2.5 \log_{10}(F) + \mathrm{ZP},
\)
where $F$ is the \texttt{PDCSAP} flux in electrons per second and $\mathrm{ZP} = 20.44$ (uncertainty $\approx 0.05$ mag). We note that, due to TESS's large pixel scale ($\approx 21^{\prime\prime}$), flux contamination from neighboring sources within the aperture may dilute the observed variability. However, the closest sources to our object are at least $25^{\prime\prime}$\ away.

\subsection{GTC EMIR JHK IR-spectroscopy}

V1082\,Sgr was observed using EMIR, a NIR imager/spectrograph \citep{Garzon_2022} on the Gran Telescopio Canarias (GTC) in queue mode (Proposal ID: GTC3-18BMEX, PI: G. Tovmassian). The observations were taken in the J, H, and K bands on 17, 20, 22, 23, and 25 September 2018 for $\sim 2770$\,s length exposures on each date with 320\,s of on-source exposure time, respectively. The exposure times were selected to evenly cover all orbital phases. The average seeing during these observations was $\sim$0.6 arcsec.
Standard procedures for data reduction and calibrations of the raw images were performed using specialized software developed by the EMIR team. For that purpose, the telluric standard star HIP92663 was also observed along the object.

\subsection{Additional data}

We also included V479\,And SDSS DR9 \& DR12 spectra for this study  \citep{2012ApJS..203...21A, 2015ApJS..219...12A}.
They were obtained with the SDSS BOSS spectrograph \citep{2013AJ....146...32S} covering the $3600-10\,400$\,\AA~wavelength range. 
Finally, we used the VizieR Catalogue to access the photometric spectral energy distribution (SED) data  \citep{10.26093/cds/vizier}.

 \section{Results}      
\subsection{\textit{TESS} light curve and ellipsoidal variability of V479 And}

\begin{figure}
\centering
\includegraphics[width=\columnwidth, bb=0 0 520 378, clip=]{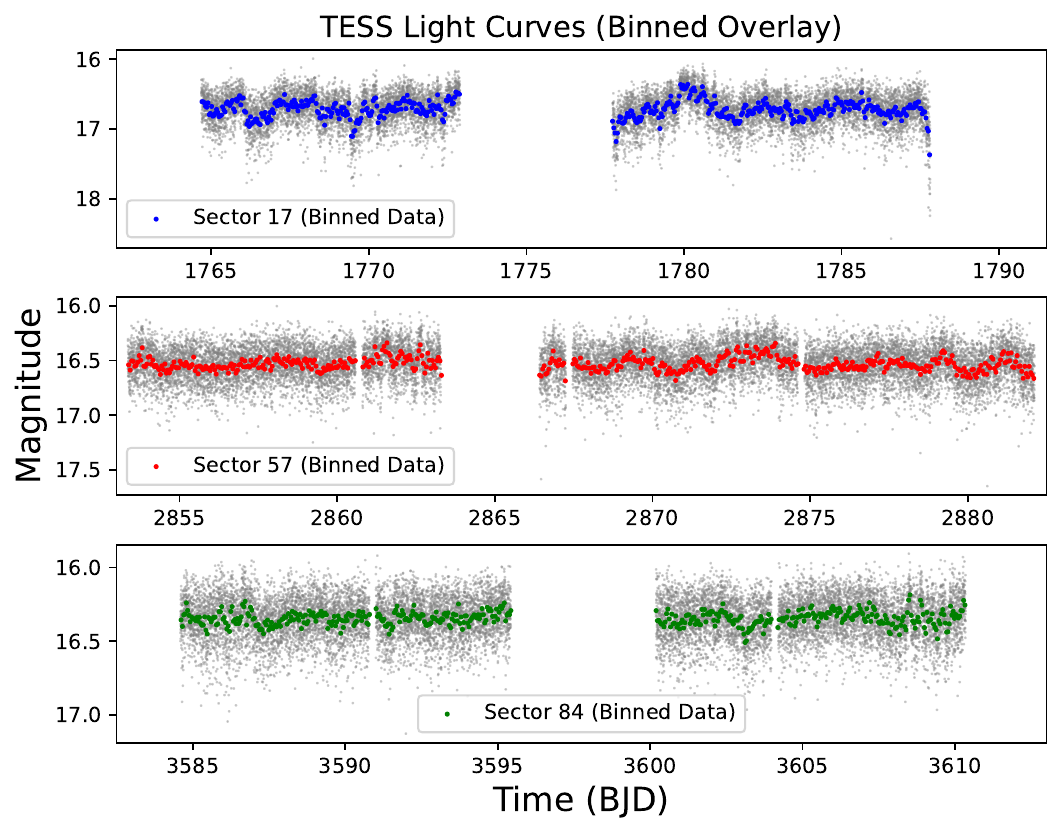}
\caption{The global TESS light curve of V479 And comprises three sectors (marked as 17, 57, and 84), each $\approx28$-day long. The original data has a cadence of 120~s. For better visualization, we binned the data to 600 points. The X-axes have similar lengths.}
\label{TESSlc}
\end{figure}

The advantage of the continuous, long monitoring provided by \textit{TESS} is that it helps to discover any small orbital modulation in the light curve of this extremely long-period CV, otherwise lost in short-episodic ground-based observations. The entire V479\,And data collected by \textit{TESS}, shown as three segments, is presented in Fig.\,\ref{TESSlc}. 

\begin{figure}
\centering
\includegraphics[width=\columnwidth, bb=0 0 300 220, clip=]{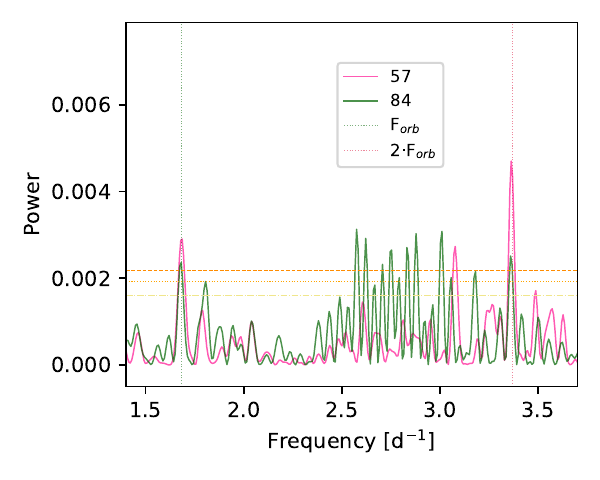}
\caption{The power spectra of V479\,And for individual sectors, i.e., 57 (pink), and 84 (green). 
The horizontal lines indicate false alarm probability levels corresponding to 0.1 (dash-dotted; yellow), 0.05 (dotted; orange), and 0.01 (dashed; dark orange). The vertical
dotted lines indicate meaningful frequencies: the red is the double orbital
frequency, the green is the orbital frequency.
}
\label{TESSpw}
\end{figure}

The period analysis was performed using the Lomb--Scargle method \citep{Lomb_1976,Scargle_1982}, as implemented in the \texttt{astropy.timeseries.LombScargle} class \citep{VanderPlas2018}. 
Fig.\,\ref{TESSpw} shows the resulting power spectra. 
The first noteworthy characteristic is that the exact orbital frequency  (F$_{\mathrm{orb}}=1.68325\,\mathrm{d}^{-1}$) corresponding to 0.594093\,d), as determined from the spectroscopy  \citep{2013A&A...553A..28G}, shows up in all sectors and is marked by a vertical green dotted line. The power is strongest in Sector 84 and weakest in 17 (the latter is very noisy and for clarity is not shown in Fig\,\ref{TESSpw}). Meanwhile, the peak at twice the orbital frequency (2$\times$F$_{\mathrm{orb}}=3.3665\,\mathrm{d}^{-1}$) is also present in all sectors, but is much stronger in Sector 57, comparable to the orbital frequency in Sectors 84 and 17.  False alarm probability (FAP) levels were derived for values of 0.1, 0.05, and 0.01, corresponding approximately to confidence levels of 90\%, 95\%, and 99\%, respectively. These FAP thresholds were computed using the analytical prescription in the Astropy implementation. All relevant power peaks reach or exceed confidence levels, plotted as horizontal lines at a false alarm probability of 0.01 for each sector\footnote{We note that these thresholds are commonly referred to as ‘False Alarm Probabilities’ in the literature \citep[e.g.,][]{press1992numerical}, although they do not strictly represent false detection rates in a Bayesian sense.}
The power of sectors 84 and 17 contains a multitude of other peaks exceeding FAP, but we did not explore them due to the low levels of variability in the data. We assume, nevertheless, that the frequencies strictly coinciding with the orbital and double orbital are not random.  

The frequency corresponding to twice the orbital frequency is commonly attributed to the ellipsoidal variability produced by the Roche-Lobe-filling donor star. The amplitude of this variability is barely 0.018 mag, preventing us from getting a meaningful phase-folded light curve. 

\subsection{HST spectrum and chemical composition of V479 And}\label{subsec:HST}

The COS spectrum of V479\,And is presented in Fig.\,\ref{HSTsp}. 
No spectral features in absorption belonging to the underlying WD are detected. The spectrum is dominated by accretion-powered emission. Such spectra 
of magnetic CVs are rather common \citep[e.g., ][]{1997ApJ...477..832M,2001ApJ...548..410S}.
They are generally interpreted as a combination of accretion-heated photospheric spots, cyclotron-emitting shock spots, or emission from the accretion stream that is responsible for the UV continuum emission \citep{1997ApJ...477..832M}. In a few cases, such as V1309 Ori, the extreme ratio of  \ion{N}{v}$\lambda1240$/{\ion{C}{iv}$\lambda1549 = 7.2$ is suggestive of an overabundance of nitrogen in the gas stream \citep{2001ApJ...548..410S}. 
The spectrum of V479\,And is also remarkable for the high intensity of \ion{N}{v} and \ion{He}{ii}$\lambda1640$ lines in comparison to very modest \ion{C}{iv}. 
The strength of the \ion{He}{ii} line confirms that V479\,And is a polar-type magnetic CV, deduced by observing repeating X-ray pulses with the orbital period \citep{2013A&A...553A..28G}. The X-ray/EUV emission from the magnetic pole provides sufficient energy to ionize \ion{He}{} in the mass transfer streams and sometimes at the irradiated face of the donor star.

The measured \ion{N}{v} / \ion{C}{iv} = 2.8 ratio (Table\,\ref{tab:UVlines}) is much larger than the value of 0.6 that is observed in the majority of CVs (e.g., the surveys of UV emission line ratios conducted by \citealt{1997ApJ...477..832M}, \citealt{Gan2003}, \citealt{GS2023}, \citealt{Sanad11}, and \citealt{2023MNRAS.523..305T}). There are currently about 30 CVs with such an inverted carbon-to-nitrogen ratio. They are all considered to have an evolved donor star that was formed from a binary containing a relatively massive secondary star with a CNO-dominated nuclear engine, which underwent an unstable thermal timescale mass transfer (TTMT) stage in its evolution \citep{10.1046/j.1365-8711.2002.05999.x,2010ApJ...717..724G,2020ApJS..249....9G,ElBad2021}. 

\begin{figure}
\centering
\includegraphics[width=\columnwidth, bb=0 0 300 218, clip=]{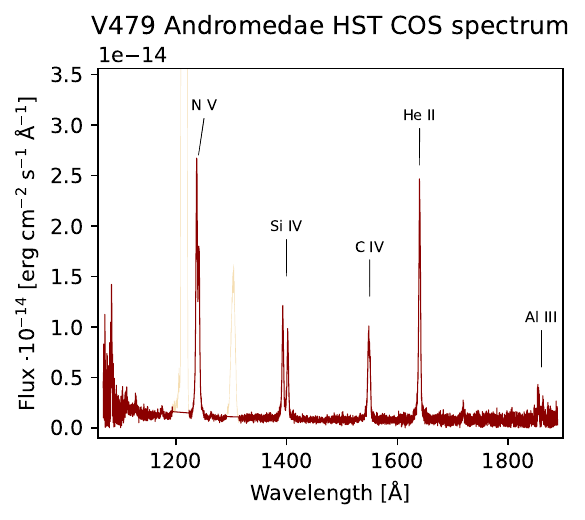}
\caption{The HST COS UV spectrum of V479\,And. The major emission lines are marked. The geocoronal lines are masked but displayed in a bleak color. }
\label{HSTsp}
\end{figure}

\begin{figure*}
\centering
\includegraphics[width=\textwidth, bb=0 0 420 160, clip=]{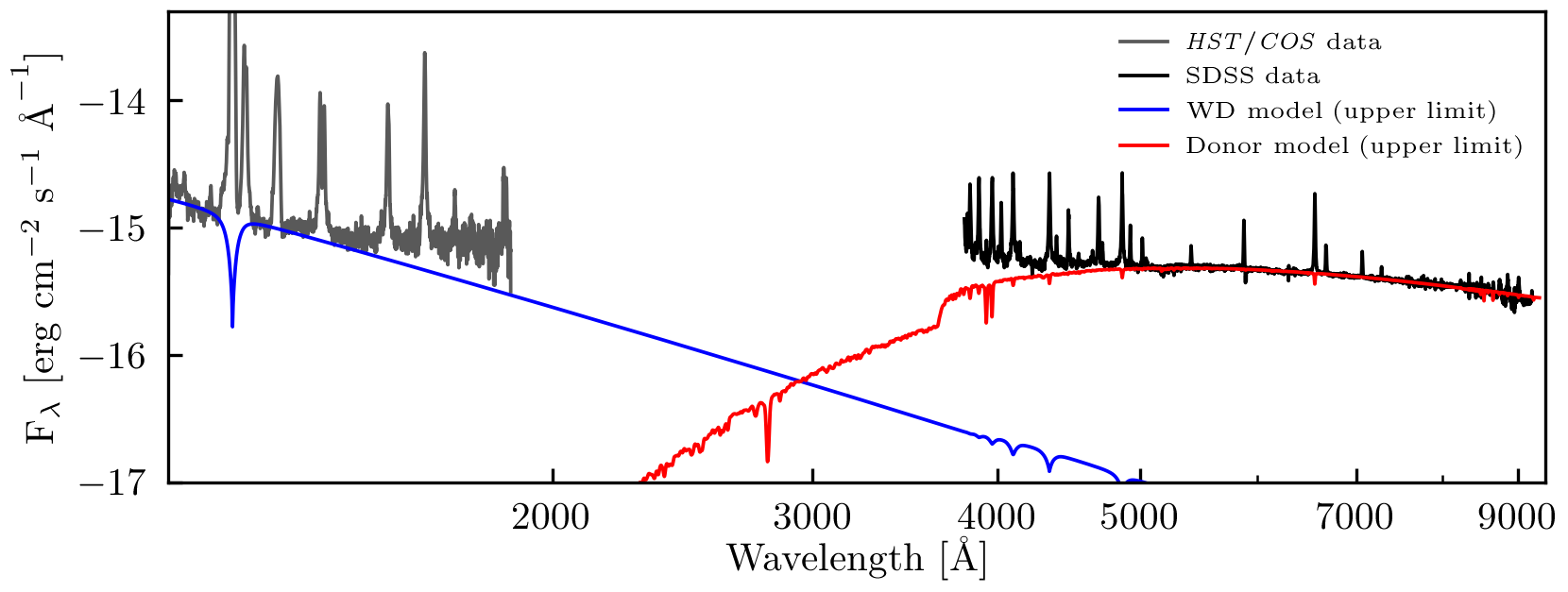}
\caption{The SED of V479\,And comprised of the {\textit {HST}}/COS ultraviolet spectrum (dark grey) and the SDSS optical spectrum (black). In blue and red are shown the upper limits for a white dwarf and a main-sequence star, respectively (see Section~\ref{subsec:HST}). The donor's model assumes a distance of 1850~pc. The white dwarf model is scaled to the same distance, and both are plotted on an absolute flux scale.
}
\label{fig:composite_sp}
\end{figure*}

The system CSS\,120422:111127+571239 is one such binary with an evolved donor, on the short extreme of the orbital period distribution.
\citet[][their fig.\,12]{2015ApJ...815..131K} illustrates perfectly how UV emission line ratios help to separate these objects from CVs formed according to a standard evolutionary scenario (i.e., those with unevolved main-sequence donor stars).

For resonance doublets such as N\,V~$\lambda\lambda1238,1242$ and Si\,IV~$\lambda\lambda1393,1402$, our COS spectra resolve the two components. We measured the flux of each component individually and then summed the fluxes to obtain the total doublet strength before computing line ratios and their logarithms. This approach ensures direct comparability with previous studies, many of which either used lower spectral resolution data or chose to report only combined doublet fluxes (e.g., \citealt{1997ApJ...477..832M, Gan2003, GS2023, Sanad11, 2023MNRAS.523..305T}). While these works generally present non-logarithmic ratios, we follow the convention of \citet{2015ApJ...815..131K} and quote the logarithm of the flux ratios in Table~\ref{tab:UVlines}, to compress the dynamic range and facilitate visual comparison with their diagnostic diagrams.

\begin{table}
    \centering
    \caption{Fluxes and logarithmic line ratios for selected emission lines in the UV spectrum of V479\,And}
    \label{tab:UVlines}
    \begin{tabular}{ccc}
    \hline \hline
      Line \AA  & Flux & {log Ratio} \\
          & $10^{-14}$ erg cm$^{-2}$ s$^{-1}$ & relative to \ion{C}{iv} \\
    \hline \noalign{\vskip 2mm}
     \ion{N}{v}   $\lambda\ {1238.82}$    & 8.56 & \multirow{2}{*}{0.45} \\
     \ion{N}{v}   $\lambda\ {1242.80}$    & 5.17 & \\
     \ion{Si}{iv} $\lambda\ {1393.75}$    & 3.65  & \multirow{2}{*}{0.13}\\
     \ion{Si}{iv} $\lambda\ {1402.77}$    & 2.95 & \\     
     \ion{C}{iv}  $\lambda\ {1550}$    & 4.92 & 0 \\
     \ion{He}{ii} $\lambda\ {1640}$    & 11.55 & 0.33 \\
    \noalign{\vskip 2mm} \hline   \hline  
    \end{tabular}
\end{table}

The composite spectrum of V479\,And, comprised of the {\textit {HST}} COS UV and the SDSS optical spectra, is shown in Fig.\,\ref{fig:composite_sp}. 
In the ultraviolet spectrum, the continuum is dominated by emission from accretion processes, which effectively ``hide'' the white dwarf. 
By this we mean that no distinct spectral features attributable to the white dwarf photosphere are detected, because they are masked by the much stronger accretion–powered continuum. 

Ignoring the dominant UV contribution from accretion, we overplot a synthetic DA model spectrum of a white dwarf with $M_1 = 0.94\,M_{\odot}$ (see Table\,\ref{Tab:FormationChannelV479And}), computed from the 3D pure–hydrogen LTE grids including H$_2$ molecular opacity \citep{Tremblay+2013,Tremblay+2015}, and scaled to a distance of 1850\,pc. 
The adopted $T_{\mathrm{eff}} = 38{,}000$\,K represents the hottest white dwarf still compatible with the short–wavelength edge of the \emph{HST} spectrum, and thus sets an upper limit to the WD temperature. 
This corresponds to the maximum temperature at which the white dwarf can remain ``hidden'' in the composite spectrum. 
Using the mass–radius relation from the La Plata group\footnote{\url{http://evolgroup.fcaglp.unlp.edu.ar/TRACKS/newtables.html}} \citep{Camisassa+2016}, these parameters yield a radius of $R_1 = 0.0088\,R_{\odot}$.

We subsequently performed a fit of the SDSS spectrum, fixing the WD mass to $0.94\,M_{\odot}$, while allowing the donor’s $T_{\mathrm{eff}}$ and mass to vary and constraining the donor radius to the Roche–lobe radius via Eq.~\ref{eq:rl_radius}. 
Using the grid of M-dwarf models by \citet{AllendePrieto+2018}, we obtained $M_{\mathrm d} = 0.24 \pm 0.10$\,M$_\odot$, $R_{\mathrm d} = 0.84 \pm 0.14$\,R$_\odot$, and $T_{\mathrm{eff}} = 5000 \pm 170$\,K. 
These values should also be regarded as upper limits, since the contribution from accretion was not included in the fit. 
The donor’s model assumes $Z=0.13$ and the same system distance of 1850\,pc. 
Importantly, the parameters derived from this spectroscopic fit are consistent with the values obtained in Sec.~\ref{sec:R2M2} from photometry and Roche–geometry arguments, reinforcing the reliability of both approaches.

\subsection{GTC IR spectrum of V1082 Sgr and chemical deviations }

\begin{figure}
\centering
\includegraphics[width=\columnwidth, bb=0 0 600 460, clip=]{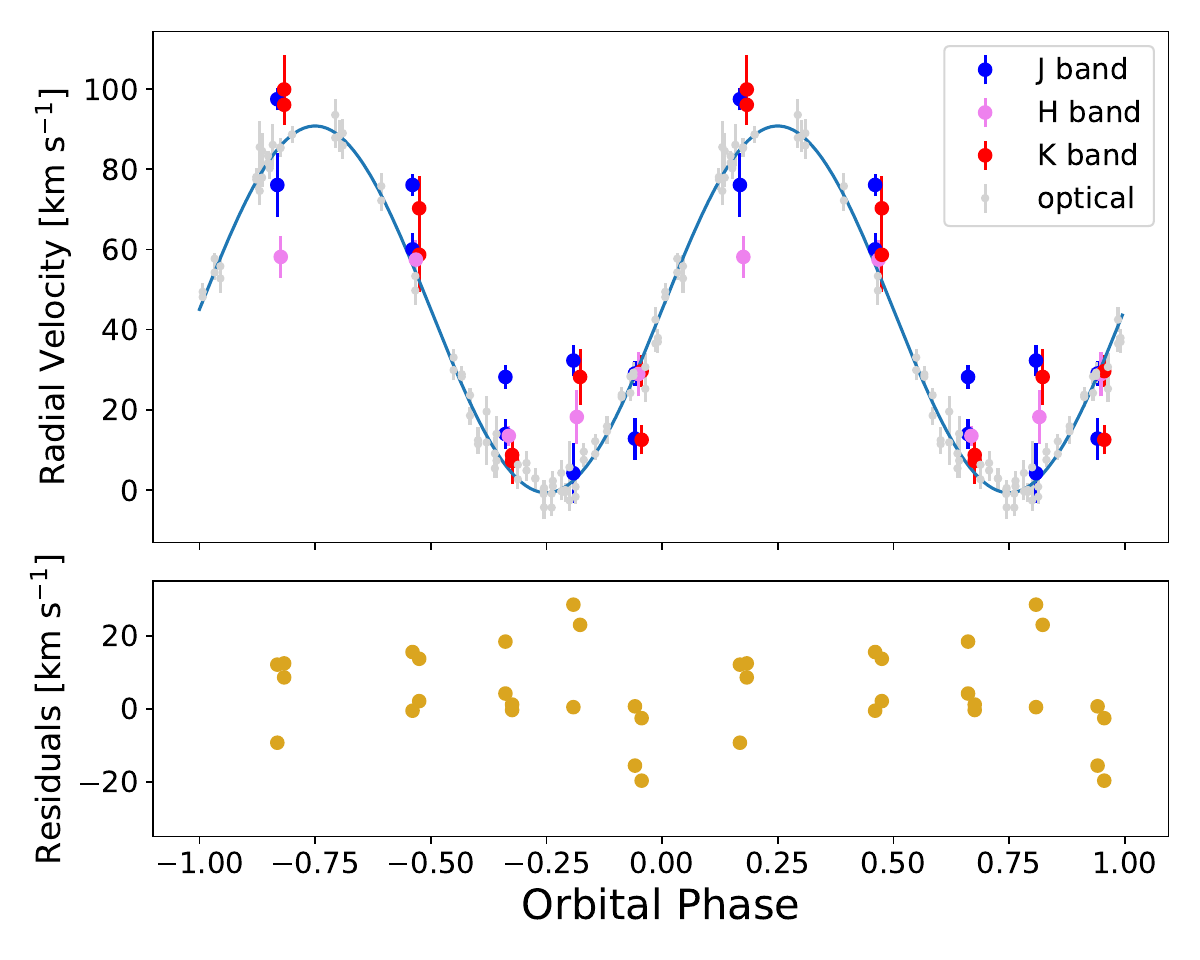}
\caption{The radial velocities measured from a complex of absorption lines in J (blue), H (magenta), K (red) bands of V1082\,Sgr spectra. The curve fits the radial velocities obtained from high-resolution optical data shown as light-gray points in the upper panel.
The residuals of J  and K measurements from the radial velocity curve obtained from optical data are presented in the bottom panel.}%
\label{fig:phase}
\end{figure}

Although we could not obtain UV spectra of V1082\,Sgr, we were able to get NIR spectra covering the full orbital cycle, revealing deviations of some key metal lines.  To constrain the spectral type and luminosity class of the donor star, we compared the observed near-infrared spectra in the $J$, $H$, and $K$ bands to a library of observed stellar spectral standards \citep{1997ApJS..111..445W,1998ApJ...508..397M,2000ApJ...535..325W} spanning spectral types K0–K5 and luminosity classes V and IV. The comparison was performed by computing the reduced chi-square statistic, defined as $\chi^2 = \sum_i \left[(F_{\mathrm{obs},i} - F_{\mathrm{std},i})/\sigma_i\right]^2$, where $F_{\mathrm{obs},i}$ and $F_{\mathrm{std},i}$ are the normalized flux densities of the observed and standard spectra, respectively, and $\sigma_i$ represents the uncertainty in the observed flux at each wavelength point. All spectra were aligned in wavelength space using cross-correlation techniques and resampled to a common grid.

In the $J$ and $H$ bands, where dwarf and K1-K1.5 subgiant standards were available, the best fit in terms of minimum $\chi^2$ was obtained consistently for subgiant stars, despite small differences in effective temperature. For the $J$ band, the minimum $\chi^2 = 0.929$ was found with a K1\,IV standard, while nearby dwarf standards such as K0\,V and K2\,V yielded slightly higher values of $\chi^2 = 1.185$ and $1.060$, respectively. In the $H$ band, a K1.5\,IV standard produced the best fit ($\chi^2 = 6.89$), compared to $\chi^2 = 8.98$ for the K2\,V dwarf. These results favor a subgiant luminosity class for the donor. In the $K$ band, where only a K0\,IV standard was available for the subgiant class, the lowest $\chi^2$ was obtained using a K3\,V dwarf template, suggesting a possible ambiguity in this band, likely due to the limited spectral coverage in available standards.

We measured radial velocities (RVs) of complexes of absorption lines in all three (J, H \& K) bands, obtaining RVs presented in Fig.\,\ref{fig:phase}. They are concurring with the much more precise RV curve from high-resolution optical observations \citep{Tovmassian_2018a}. Moreover, the \ion{Na}{i}, \ion{Mg}{i} lines, which we will consider subsequently, follow the RV curve of the donor star, that is, do not have an interstellar or other exotic origin. 
A large fraction of NIR spectra are presented in Fig.\,\ref{fig:IR}. Significant lines from \citet[][table\,7]{2009ApJS..185..289R} are marked, where the spectra of the corresponding standards are from the databases described in \citet{1997ApJS..111..445W}, \citet{1998ApJ...508..397M}, and \citet{2000ApJ...535..325W}.

\begin{figure*}[t]
\setlength{\unitlength}{1mm}
\resizebox{9.2cm}{!}{
\begin{picture}(90,60)
\put (0,10){\includegraphics[width=5.95cm, bb=0 40 480 420, clip=]{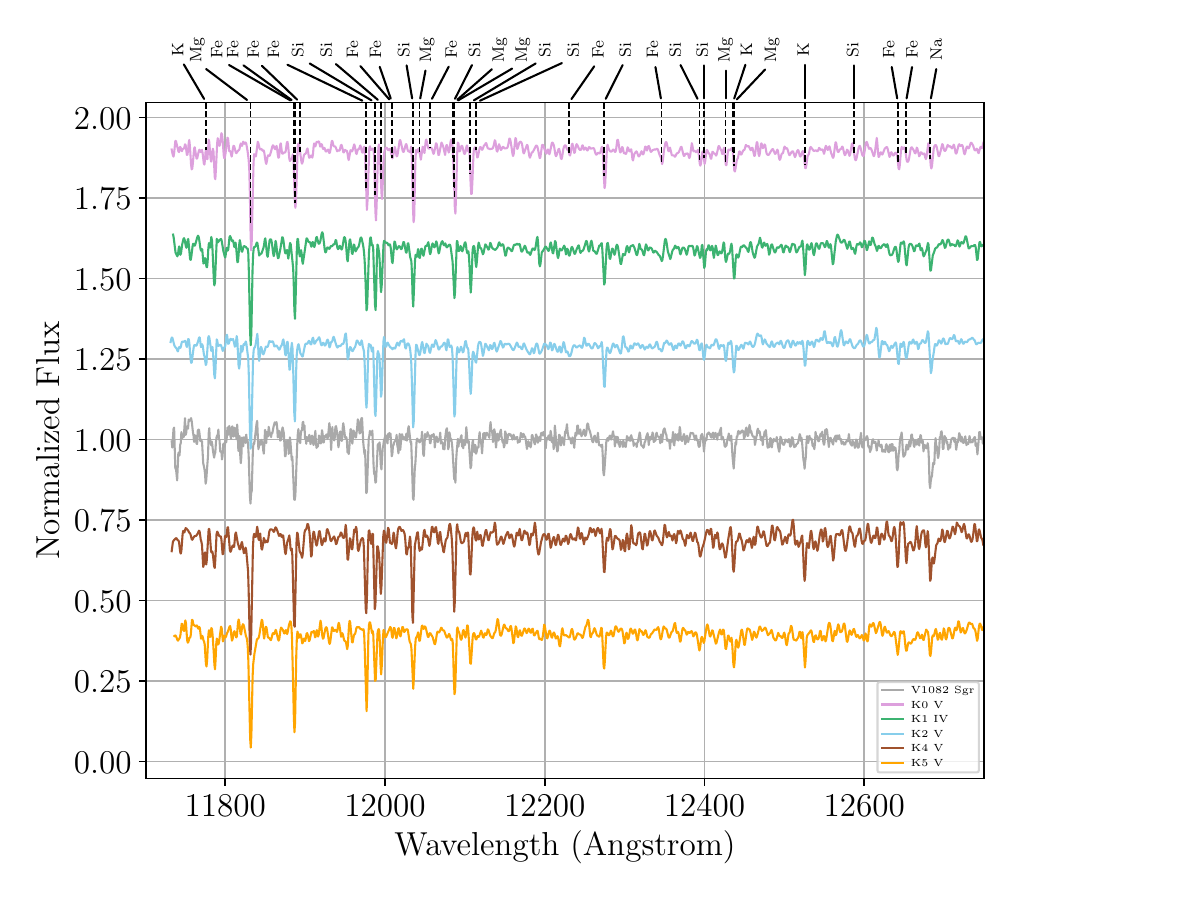}}
\put(60,5){\includegraphics[width=5.6cm, bb=30 0 480 420, clip=]{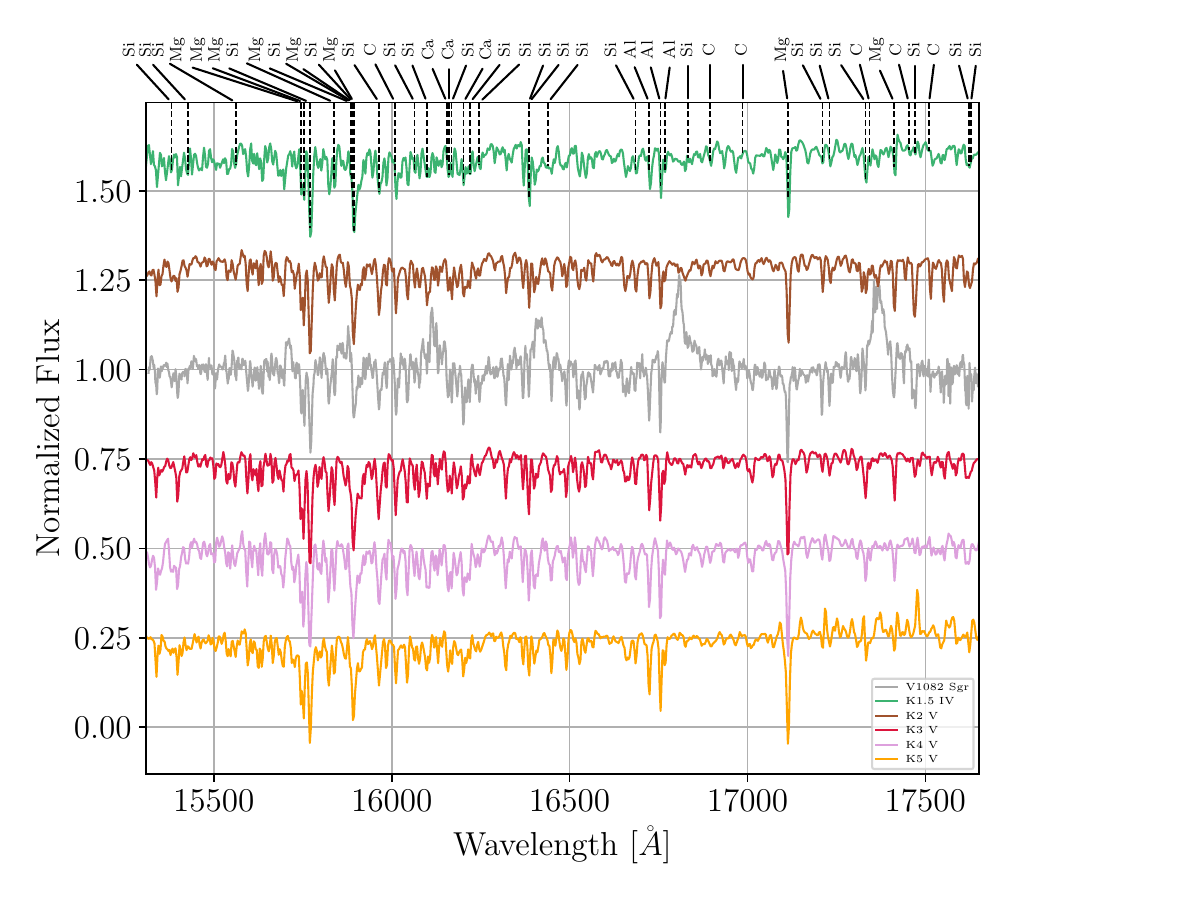}}
\put(117,10){\includegraphics[width=5.6cm, bb=30 40 480 420, clip=]{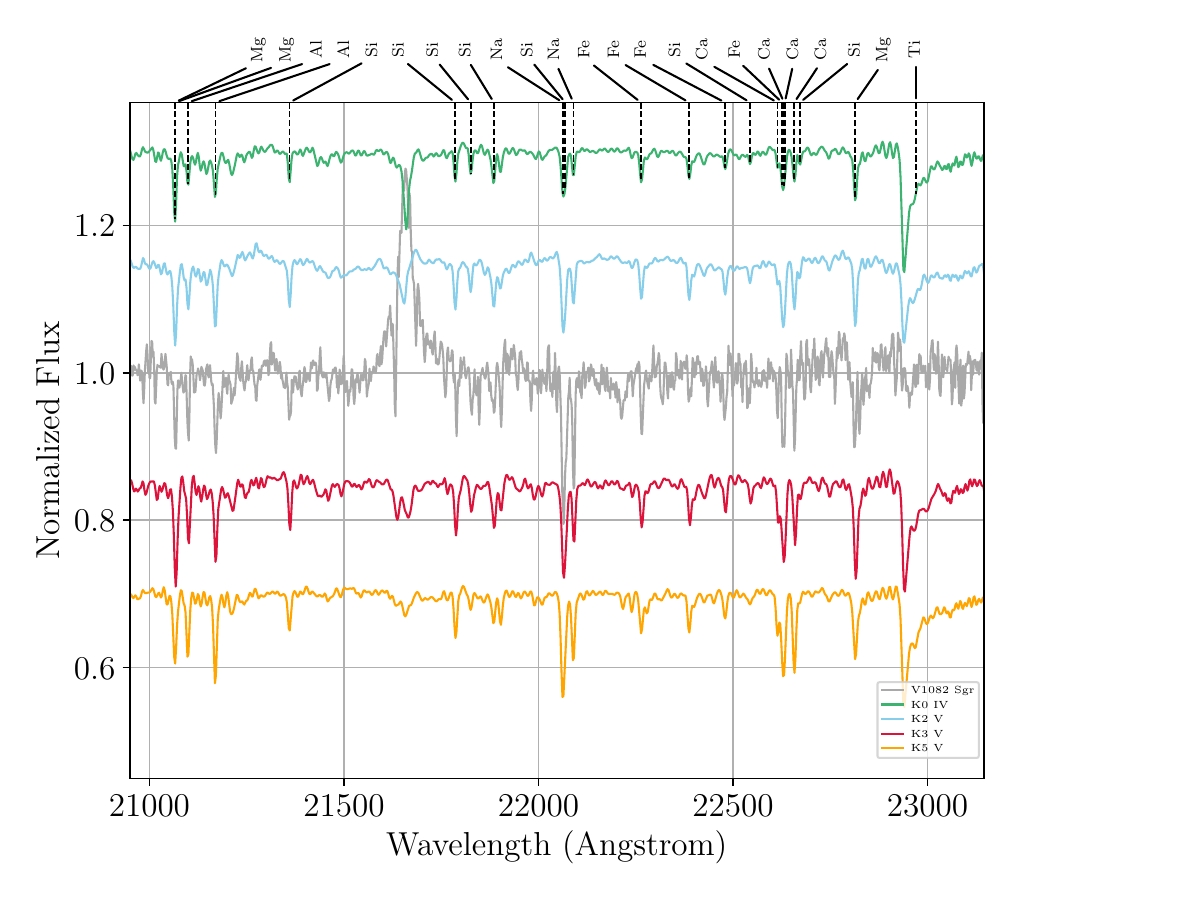}}
\end{picture}}
\caption{The IR spectra of V1082\,Sgr. The spectra are co-added from five individual spectra taken evenly around orbital phases. They were shifted to zero velocity after measuring RVs before co-adding and normalizing. They are placed between K2 and later spectral-type standards because they best resemble the K2 spectral class, or slightly earlier if compared to IV luminosity class.}
\label{fig:IR}
\end{figure*}

\begin{figure*}[t]
\centering
\includegraphics[width=8.3cm, bb=0 0 600 480, clip=]{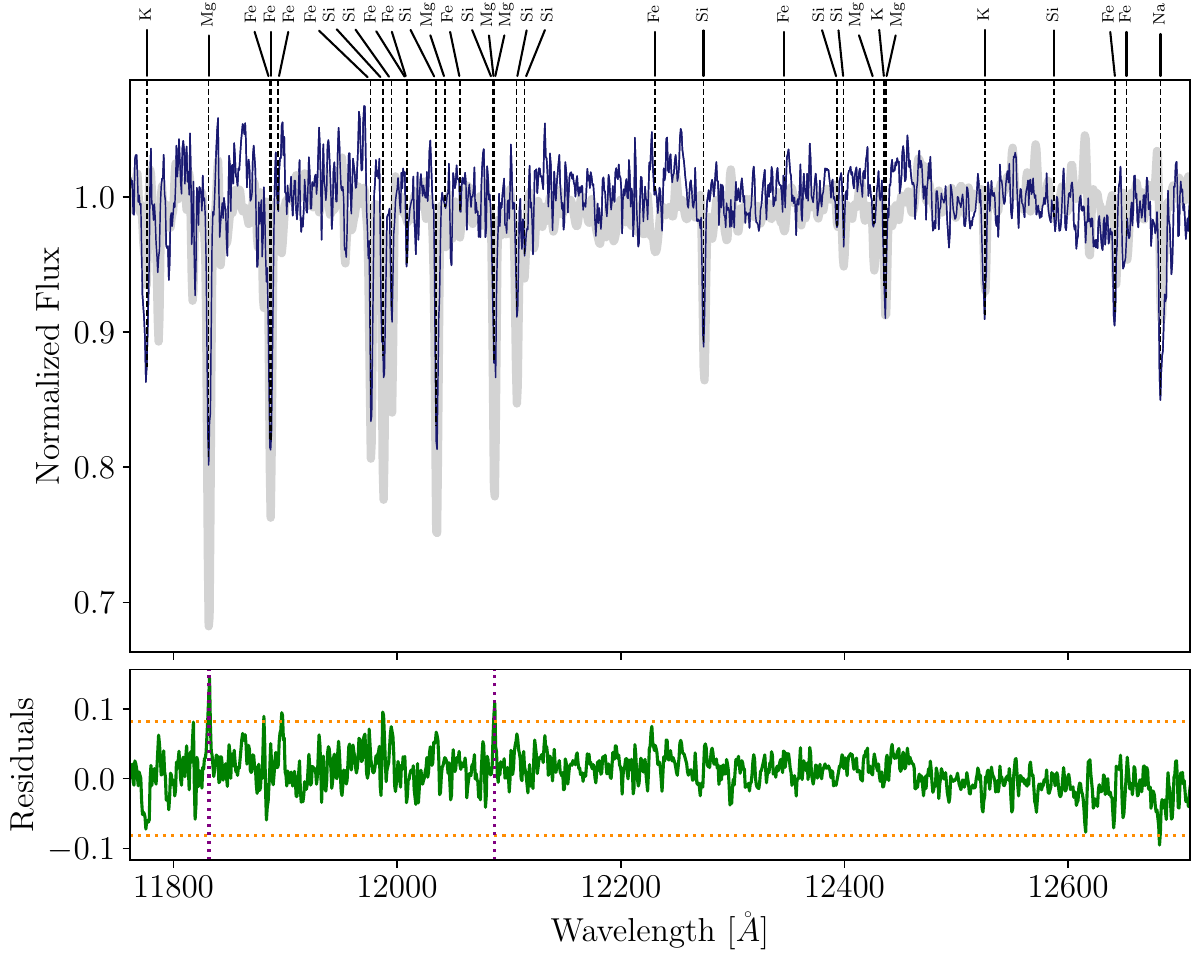}
\includegraphics[width=8cm, bb=22 0 600 480, clip=]{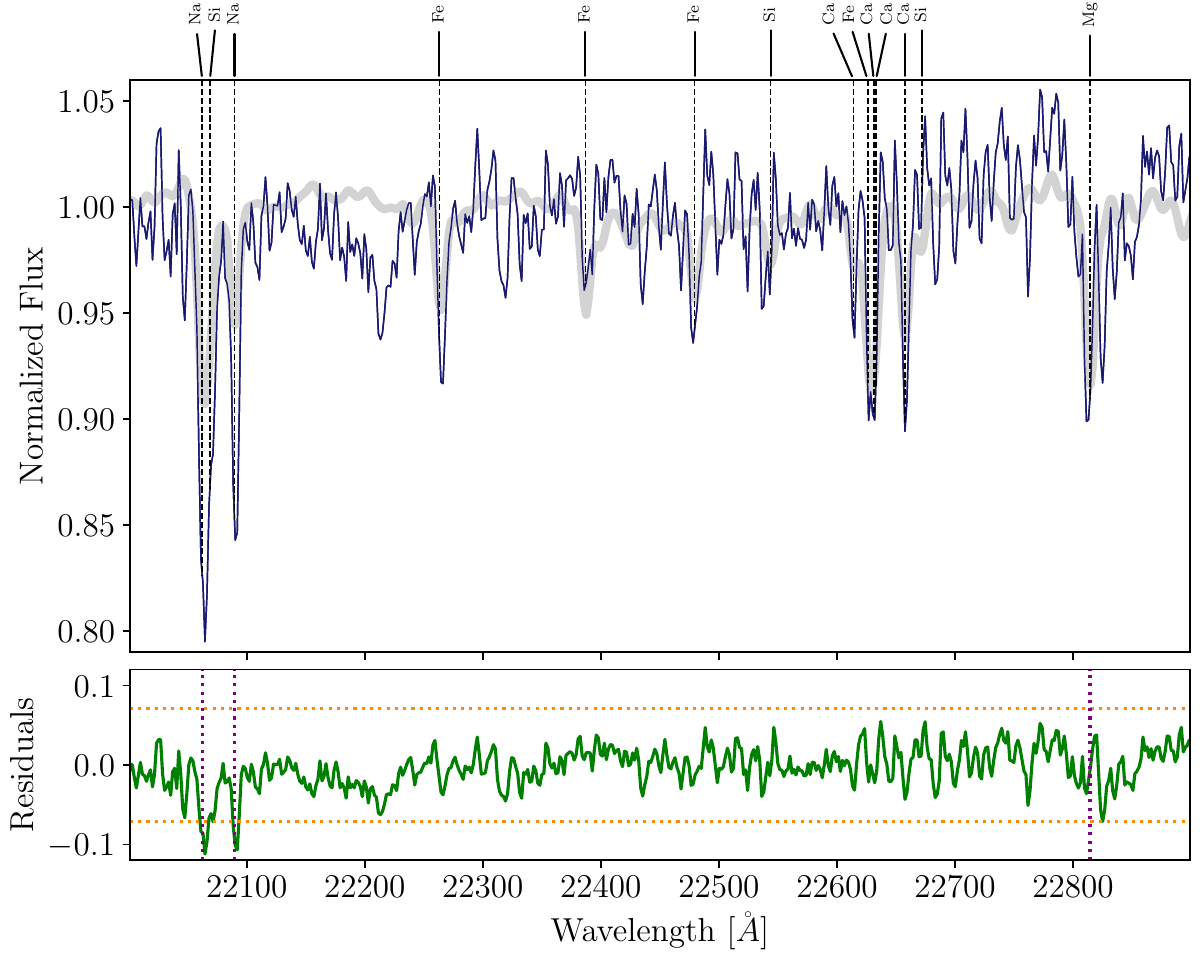}
\caption{Portions of the IR spectra of V1082\,Sgr, emphasizing spectral lines in discordance with the corresponding standards, are presented in the upper panel. The residuals between the observed normalized spectrum and a K2\,V type standard are shown in the bottom panels. Horizontal dotted orange lines mark the $3\sigma$ deviation level. Vertical dotted lines mark the deviating lines.}
\label{fig:JKIR}
\end{figure*}

The doublets \ion{Na}{i} lines at $\lambda\ 22062.4\ \&\ 22089.7\ \AA$ show orbital variability, testifying that they originate in the atmosphere of the donor star. The zoom-in portions of the co-added and normalized spectra are shown in Fig.\,\ref{fig:JKIR} together with standard stars, testifying that despite good accordance, there is a significant excess \ion{Na}{i} and deficit of \ion{Mg}{i} in the spectrum of V1082\,Sgr. \ion{Na}{i} $\lambda\ 22062.4\ \AA$\ has an equivalent width EW$=2.65\pm0.05$\,$\AA$, while it is expected for the K2 spectral type to be around 1.66  \citep{2009ApJS..185..289R}.

The presence of an enhanced \ion{Na}{i} doublet and a deficit of \ion{Mg}{i} in V1082 Sgr also aligns with the findings of \citet{2018ApJ...861..102H}, who noted that these chemical anomalies are common in hydrogen-deficient CV donor stars, particularly those with evolved secondaries.  \citet[][their fig.\,1]{2018ApJ...861..102H} provided excellent evidence of an increasing strength of \ion{Na}{i} doublet toward the hydrogen-poor synthetic spectra. The enhanced sodium abundance in V1082\,Sgr is particularly striking, suggesting a donor star that has undergone substantial nuclear evolution, possibly due to its higher initial mass compared to typical CV donors.

The comparison with the CVs mentioned above, which harbor evolved secondaries, is telling. These systems exhibit evolved donor stars but do not show the same substantial sodium excess observed in V1082\,Sgr. The absence of a similar \ion{Na}{i} enhancement in these systems, despite their evolved nature, highlights the unusual and distinctive nature of the V1082\,Sgr donor star. This divergence in chemical behavior suggests that the donor star in V1082\,Sgr may have evolved through a different, possibly more extreme, evolutionary path, which could be linked to its higher initial mass or alternative evolutionary mechanisms, such as TTMT. Therefore, the sodium excess in V1082 Sgr strengthens the argument that its donor is not just evolved but may represent an alternative evolutionary pathway for CV evolution.

\begin{figure*}[t]
\centering
\includegraphics[width=8cm, bb=0 0 560 410, clip=]{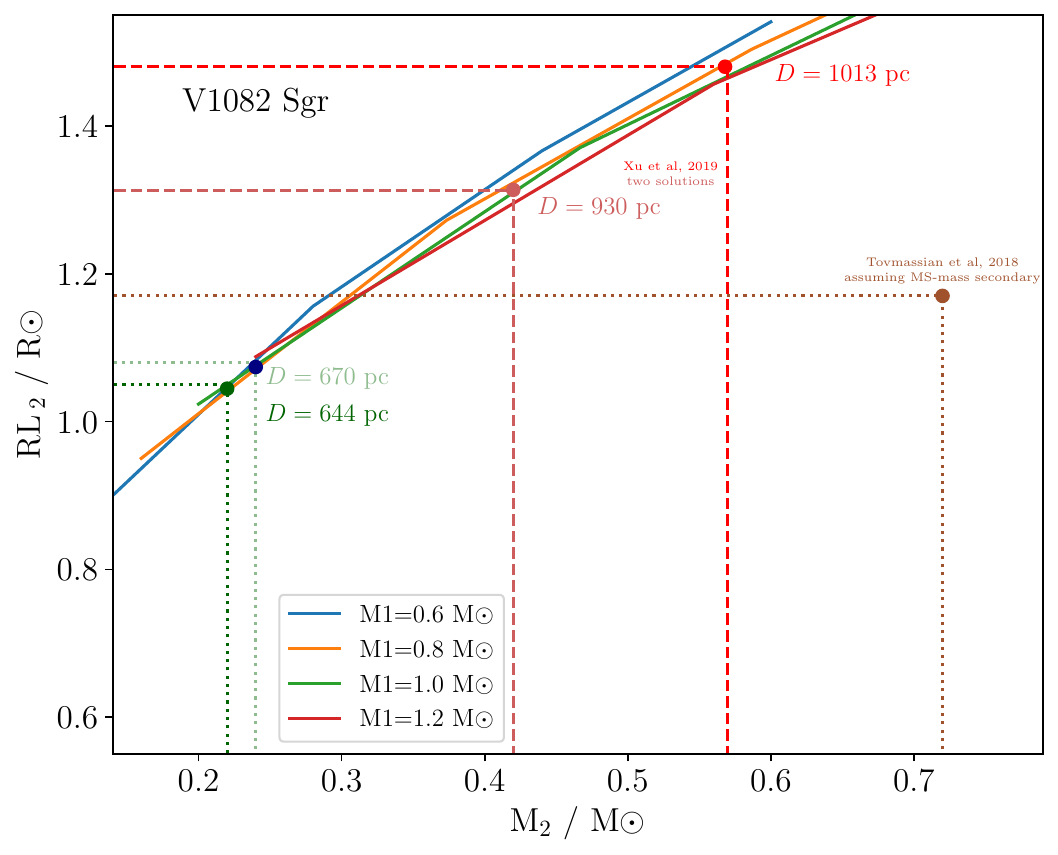}
\includegraphics[width=8cm, bb=5 0 560 410, clip=]{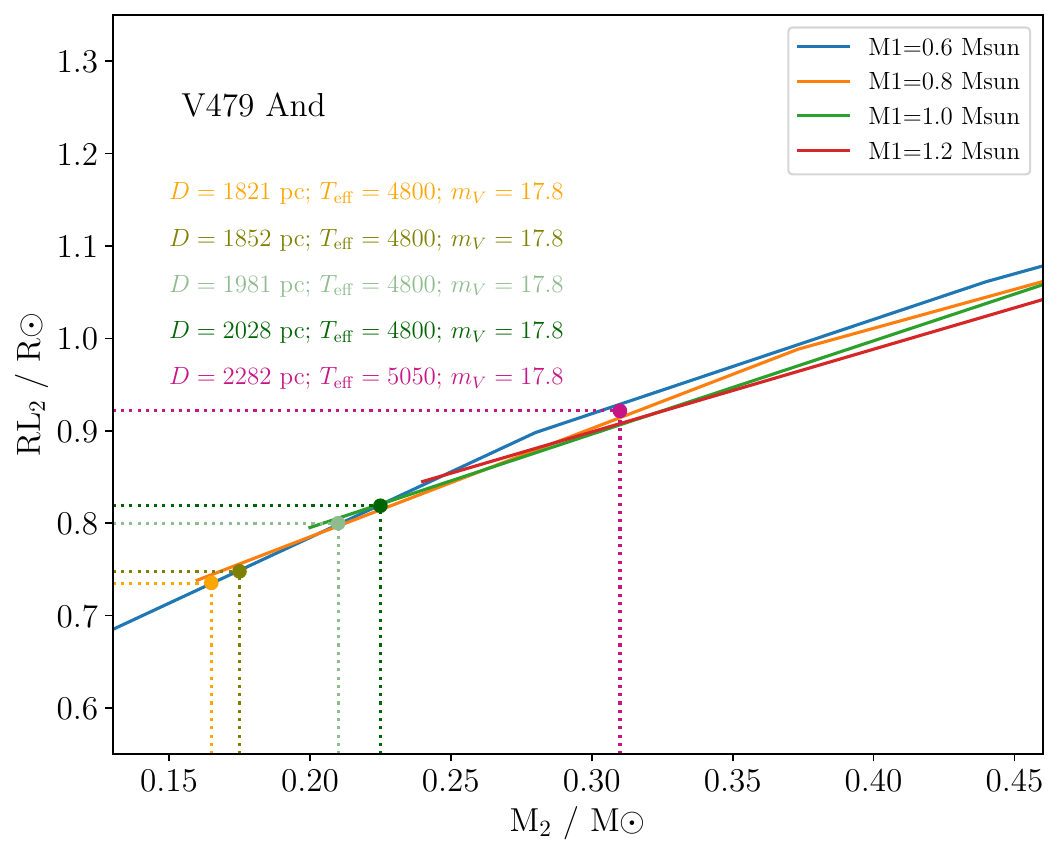}
\caption{Diagnostic diagrams of the $R_2$–$M_2$ relation for V1082~Sgr (left) and V479~And (right), assuming that the donor star fills its Roche lobe and accounting for the revised Gaia distance ranges. The methodology is described in Sect.~4.1. Horizontal dotted lines indicate the donor star radii (assuming spherical geometry) corresponding to the spectral class $T_{\mathrm{eff}}$ needed to match the observed luminosity for different distances. Vertical lines mark the inferred donor masses. In the left panel, the solutions fall within the shaded green area. In contrast, solutions from the previous studies (labeled) illustrate that they are inconsistent with the revised distances adopted in this work.}
\label{fig:r2m2}
\end{figure*}

\section{Basic parameters of the donor stars}
\label{sec:masses}

%
We adopted improved \gaia\ DR3 distances in our following analyses.
With those, we can derive the parameters of the donor stars in \VSgr~and \VAnd.

\subsection{Radii and masses}
\label{sec:R2M2}

The mass of the donor star in a close binary system containing a WD can be estimated using a combination of Roche geometry, Kepler's third law, and distance measurements. In systems where the donor star fills its Roche lobe and undergoes mass transfer, the size of the Roche lobe can be related to the mass ratio of the system, providing a means to estimate the donor's mass. Here, we outline a methodology to estimate the donor star's mass by constructing a mass--radius relation, leveraging the Eggleton approximation for the Roche lobe, and incorporating precise distances from the \emph{Gaia} mission to derive absolute magnitudes and luminosities.

We use the \emph{volume--equivalent Roche--lobe radius} $R_{\mathrm{L,ve}}$\footnote{We will refer to $R_{\mathrm{L,ve}}$ simply as the Roche lobe radius $R_{\mathrm{L}}$ further in the text}, 
i.e. the radius of a sphere having the same volume as the Roche lobe, 
computed with the Eggleton approximation \citep{eggleton1983}:
\begin{equation}\label{eq:rl_radius}
    \frac{R_L}{a} = \frac{0.49 q^{2/3}}{0.6 q^{2/3} + \ln(1 + q^{1/3})},
\end{equation}
\noindent
where \( a \) is the binary separation and \( q = \frac{M_2}{M_1} \) is the mass ratio of the donor star to the WD. Assuming that the donor star fills its Roche lobe, we have \( R_2 \approx R_L \), where $R_2$ is the donor radius.

The orbital separation \( a \) can be derived from Kepler’s third law as
\begin{equation}
    a^3 = \frac{G (M_1 + M_2) P_{\rm orb}^2}{4 \pi^2},
\end{equation}
\noindent
where \( G \) is the gravitational constant, \( M_1 \) is the mass of the WD, \( M_2 \) is the mass of the donor star, and \( P_{\rm orb} \) is the observed orbital period. From this, the orbital separation \( a \) can be computed for a given system.

To estimate the donor star’s mass, we construct a mass-radius relation for a range of WD masses. For each assumed \( M_1 \), we calculate the Roche lobe radius \( R_L \) as a function of \( M_2 \), using the Eggleton formula and the corresponding binary separation \( a \) from Kepler's law. This generates a set of curves in the \( R_2 \)-\( M_2 \) plane for various assumed WD masses.

At the same time, the radius \( R_2 \) of the donor star can be independently determined from its observed photometric properties. The donor star’s luminosity \( L_2 \), radius \( R_2 \), and effective temperature \( T_2 \) are related by the dimensionless relation:

\begin{equation}
    \frac{L_2}{L_\odot} = \left( \frac{R_2}{R_\odot} \right)^2 \left( \frac{T_2}{T_\odot} \right)^4,
\end{equation}
\noindent
where \( L_\odot \), \( R_\odot \), and \( T_\odot \) are the solar luminosity, radius, and temperature, respectively. This allows us to estimate the radius \( R_2 \), based on the effective temperature \( T_2 \) derived from the spectrophotometric energy distribution fitting.

The luminosity \(L\) is derived from the absolute bolometric magnitude \(M_{\rm bol}\), which we compute from the \emph{observed} \(V\)-band apparent magnitude, the distance, the line-of-sight extinction, and the bolometric correction. First, the absolute visual magnitude is
\begin{equation}
    M_V = m_V - 5\log_{10}(d) + 5 - A_V,
\end{equation}
\noindent
where \(m_V\) is the observed apparent magnitude (not de-reddened), \(d\) is the distance in parsecs, and \(A_V\) is the extinction in the \(V\) band. The bolometric magnitude is then
\begin{equation}
    M_{\rm bol} = M_V + BC,
\end{equation}

In revisiting the distances adopted for the donor star analysis, we corrected the Gaia DR3 parallaxes for the zero point and recalculated the distances using both geometric and photogeometric estimates from \citep{2021AJ....161..147B}. For V479~And, the corrected parallax yields a distance of $1981$~pc ($\varpi = 0.505$~mas), while the Bailer-Jones \citep{2021yCat.1352....0B} estimates provide $2028^{+407}_{-259}$~pc (geometric) and $2282^{+527}_{-253}$~pc (photogeometric). Applying an exponentially decreasing space density prior with a scale height of 250~pc, more appropriate for CVs above the period gap, we obtain a refined distance of $1852^{+253}_{-203}$~pc. This is consistent with the \textit{Gaia} DR3 spectrophotometric estimate ($\mathrm{distance\_gspphot} = 1821$~pc) and within the error range given the moderate \texttt{ruwe} of 1.0. Similarly, for V1082~Sgr, the corrected parallax ($\varpi = 1.55$~mas) gives $644$~pc, fully consistent with the Bailer-Jones estimate ($644 \pm 7$~pc geometric, $643^{+7}_{-6}$~pc photogeometric) and our prior-adjusted distance of $644 \pm 8$~pc. The radii-mass diagnostic diagram (Fig.~8) shows a range of solutions that account for the uncertainty in distance determination.


We used bolometric corrections corresponding to the spectral classes of the donor stars, adopting the tables of \citet{1996ApJ...469..355F} and \citet{2025arXiv250319965E}. There is evidence that the bolometric corrections for K-type subgiants do not differ significantly from those of dwarfs; on the other hand, the choice of method for calculating BC introduces a larger source of uncertainty \citep{Worthey_2011,2025PARep...3...43E}.

For V1082\,Sgr, we adopted an apparent visual magnitude of $m_V \simeq 14.8$ at minimum brightness, as measured by \citet{Tovmassian_2016}. For V479\,And, no reliable direct photometric measurement of the donor exists. Instead, we used the same dataset as \citet{2013A&A...553A..28G}, but estimated the donor’s contribution by scaling and subtracting template spectra of late-type standard stars until the absorption lines were optimally balanced and the continuum was smooth. This spectral-decomposition procedure yields an effective donor brightness equivalent to $m_V = 17.8 \pm 0.1$. These values should be regarded as \emph{upper limits} to the donor brightness, since residual accretion light or the white dwarf may still contribute even in low states.

The intersection of the calculated donor radius \( R_2 \) and the Roche lobe curves in the \( R_2 \)-\( M_2 \) provides a solution for \( M_2 \), the mass of the donor star.
In Fig.\,\ref{fig:r2m2}, we present the Eggleton approximation for Roche lobe size for various \( M_1 \) as a series of curves. The horizontal lines indicate possible \( R_2 \) derived from the estimated effective \( T_2 \), \( m_{\mathrm{V}} \) and the distance. 
We estimated the \( M_2 \) range, assuming that the donor star fills its Roche lobe, and using the observed photometric brightness in the $V$ band \citep{2013A&A...553A..28G,Tovmassian_2018a} of these two systems. However, the distance,  the effective temperature, and even the brightness of the donor star are known to have large uncertainties. Hence, we have a series of \( R_2 \) values that depend on the adopted observables. The vertical lines indicate \( M_2 \) for different sets of parameters. While it is difficult to pinpoint the exact mass of the donor star, we can state unambiguously that the donor star is drastically smaller than an isolated main-sequence or pre-main-sequence star of corresponding temperature.

In the case of V1082\,Sgr, we know well the brightness of the donor star, since the object often goes to very low states, in which the naked donor star is observed. The distance error is also relatively small. We get an M$_2 \approx 0.22-0.24$\,\Msun~estimate for the donor mass (blue dots and dotted lines). Also shown are other solutions; the green point and dotted lines correspond to the proposed model with an unevolved, under-filling its Roche lobe donor star, and for \( R_2 \)'s obtained by \citet{2019MNRAS.489.3031X} models. They will require much larger distances to the object than measured by \gaia.

There are no robust dynamical constraints on the masses of white dwarfs. For V1082\,Sgr,   indirect estimates from X-ray studies exist. \citet{Bernardini_2013} 
derived $M_1 \simeq 0.64\,M_\odot$ from the X-ray luminosity, while 
\citet{2019MNRAS.489.3031X} obtained a higher value of $M_1 \simeq 0.77\,M_\odot$ by 
fitting the X-ray spectrum. Such differences are typical for X-ray based mass 
determinations in intermediate polars, which suffer from systematic uncertainties 
(e.g., \citealt{2025arXiv250603711S}) related to assumptions about accretion column 
structure, shock height, and partial covering absorption. We therefore consider 
these values as indicative only, rather than precise dynamical measurements.  

For V479\,And, the distance limits are wider, and the donor star’s temperature and 
brightness are deduced from the SED with significant uncertainties. All possible 
solutions, however, correspond to a donor mass of $M_2 < 0.31$\,M$_\odot$.  
In both systems, the adopted $M_{\rm WD}$ has no significant influence on our donor 
mass estimates. The ranges in $T_{\rm eff}$ and $m_V$ of the secondary star are 
much larger than the modest differences introduced by the white dwarf mass.

\subsection{Orbital Inclination}
\label{sec:inc}

The orbital inclination can, in principle, be derived from the observed mass function,
\begin{equation}
f(M) = \frac{P_{\rm orb} K_2^3}{2 \pi G} = \frac{(M_1 \sin i)^3}{(M_1 + M_2)^2},
\end{equation}
where $K_2$ is the radial velocity semi-amplitude of the donor star. Observations fix the left-hand side, but the inclination angle $i$ can only be determined if both stellar masses are specified. 

Although one could, in principle, insert plausible average values of $M_1$ and $M_2$, there are no robust dynamical constraints on either system. We therefore adopted the values of $M_1$ and $M_2$ provided by our evolutionary models (see Sec.~\ref{sec:diogo}) to evaluate $i$. This exercise shows that both systems must be observed at relatively low inclinations ($i \simeq 13^\circ$--$18^\circ$). Moreover, V1082\,Sgr is inferred to be slightly less inclined than V479\,And.

While such low inclinations are a priori unlikely ($\sim3\%$ for a single system and $\sim0.1\%$ for two), the observational constraints leave no alternative: both systems are best explained by low-inclination geometries.}


\begin{figure*}
\begin{center}
\includegraphics[width=0.85\linewidth]{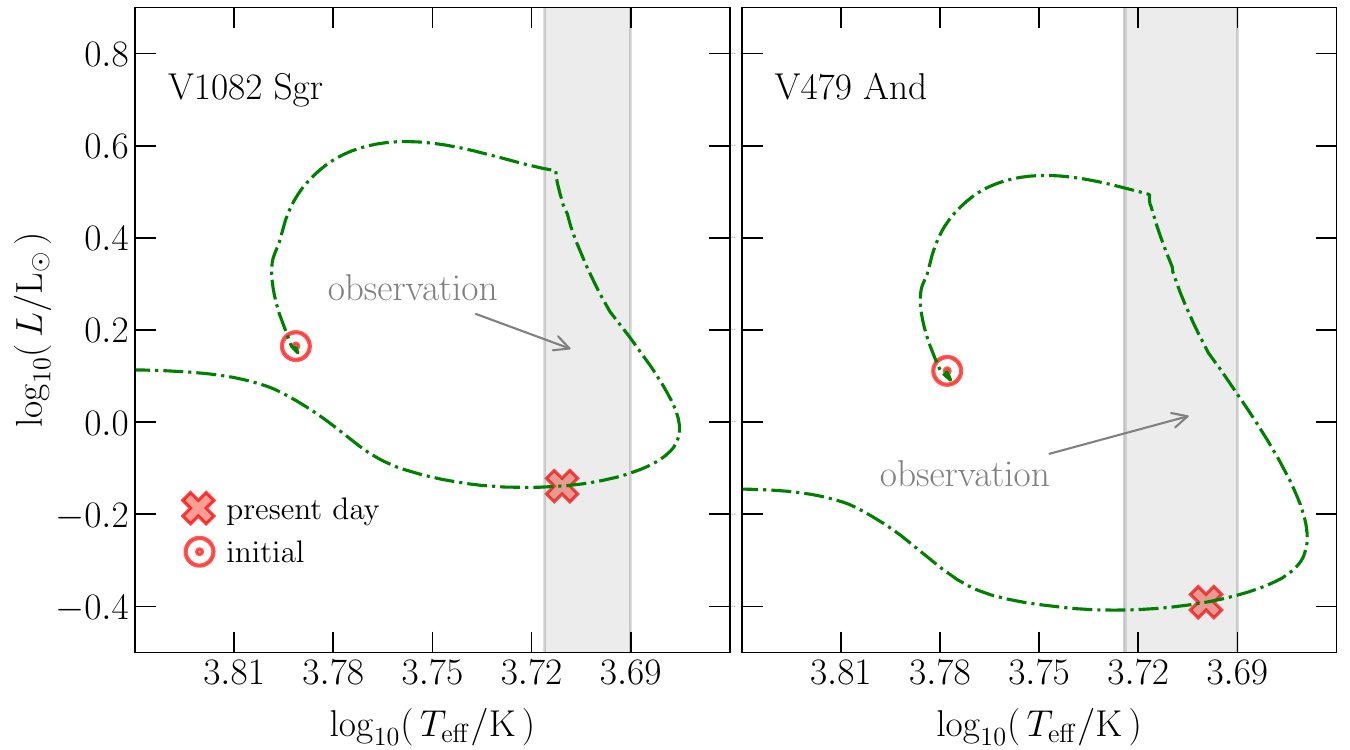}
\end{center}
\caption{Post-CE evolution of the luminosity with the effective temperature of the donor in \VSgr~(left panel) and \VAnd~(right panel). The red solar symbol indicates the properties at the moment the WD was formed (i.e., just after CE evolution), and the thick red cross indicates the properties at the moment the orbital period is the same as observed. The thick grey vertical strip corresponds to the properties derived from observations. More details are provided in Sect.~\ref{sec:diogo}.}
\label{Fig:LvsT}
\end{figure*}

\begin{figure*}
\begin{center}
\includegraphics[width=0.85\linewidth]{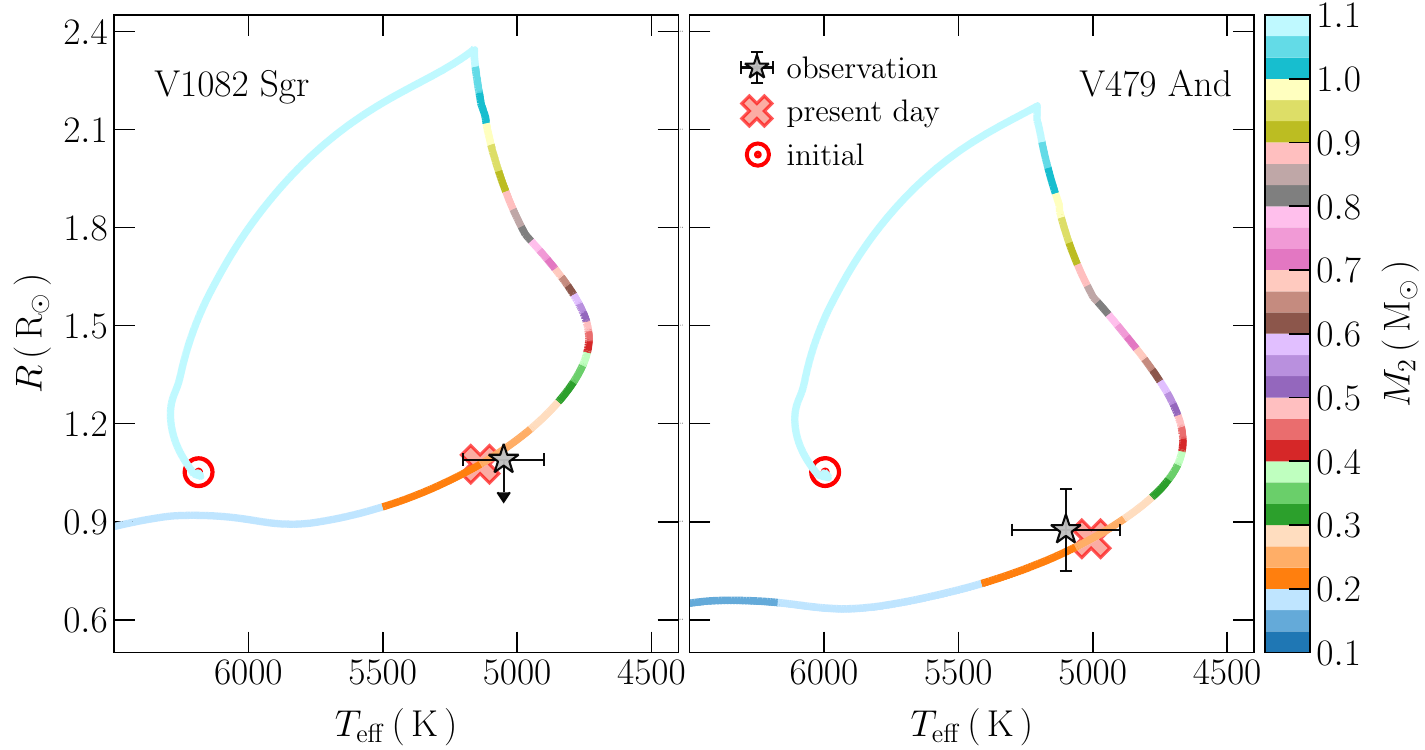}
\end{center}
\caption{Post-CE evolution of the radius with the effective temperature of the donor in \VSgr~(left panel) and \VAnd~(right panel), colour-coded by its mass. The red solar symbol indicates the properties at the moment the WD was formed (i.e. just after CE evolution) and the thick red cross the properties at the moment the orbital period is the same as observed. The grey star corresponds to the properties derived from observations. More details are provided in Sect.~\ref{sec:diogo}.}
\label{Fig:RvsT}
\end{figure*}

\begin{figure*}
\begin{center}
\includegraphics[width=0.85\linewidth]{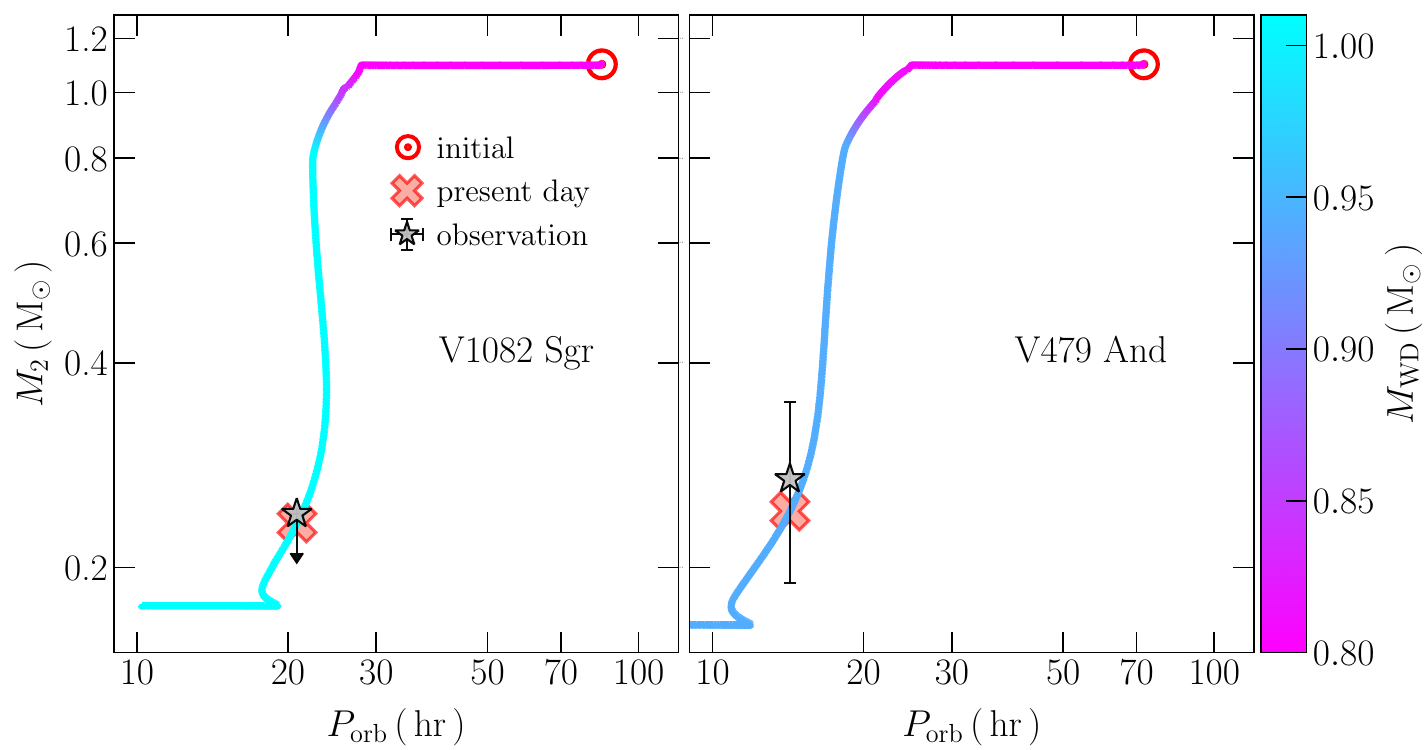}
\end{center}
\caption{Post-CE evolution of the donor mass with the orbital period, colour-coded by the WD mass, for \VSgr~(left panel) and \VAnd~(right panel). The red solar symbol indicates the moment at which the white dwarf was formed, while the thick red cross the properties at the moment the orbital period is the same as observed. More details are provided in Sect.~\ref{sec:diogo}.}
\label{Fig:PvsM}
\end{figure*}

\begin{figure*}
\begin{center}
\includegraphics[width=0.85\linewidth]{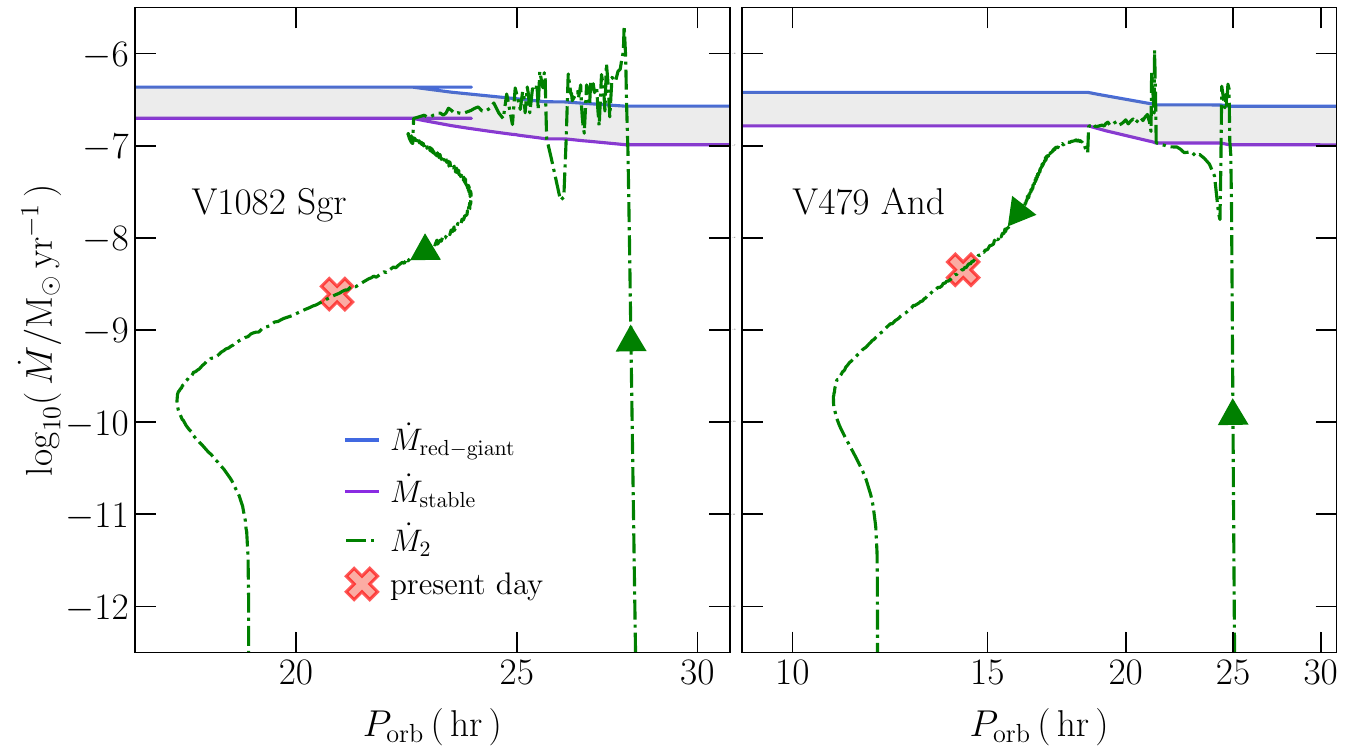}
\end{center}
\caption{Post-CE evolution of the mass transfer rate (green dot-dashed line) with the orbital period for \VSgr~(left panel) and \VAnd~(right panel). The thick red cross indicates the properties at the moment the orbital period is the same as the observed. The purple line corresponds to the critical mass transfer rate $\dot{M}_{\rm stable}$ above which hydrogen can burn steadily in a shell. The blue line corresponds to the maximum burning rate $\dot{M}_{\rm red-giant}$ such that, for higher mass transfer rates, not all the transferred mass is accreted. The non-accreted mass piles up above the WD and is eventually lost from its vicinity in the form of fast winds similar to those of red giants. More details are provided in Sect.~\ref{sec:diogo}.}
\label{Fig:PvsMdot}
\end{figure*}

\section{Evolutionary models}
\label{sec:diogo}

We can use the parameters of the donors in \VAnd~and \VSgr~we derived in the previous sections to search for decent evolutionary models that can explain both extremely long-period CVs.
We used the MESA code \citep[ version r15140][]{Paxton2011,Paxton2013,Paxton2015,Paxton2018,Paxton2019,Jermyn2023} for that and our assumptions for single star and binary evolution are described in the Appendix~\ref{app:ModelingMESA}.
We were able to successfully find reasonable formation pathways for both \VAnd~and \VSgr, which are provided in the Appendix~\ref{app:FormationChannels} (Tables~\ref{Tab:FormationChannelV1082Sgr} and \ref{Tab:FormationChannelV479And}).
The properties we predict in our modeling with MESA are compared to the observed in Table~\ref{Tab:Comparison}.

\begin{table}
\caption{Predicted and observed parameters of V1082\,Sgr and V479\,And.}
\label{Tab:Comparison}
\centering
\setlength\tabcolsep{4.5pt} 
\renewcommand{\arraystretch}{1.45} 
\begin{tabular}{lcccc}
\hline\hline
\multirow{2}{*}{Parameter} & \multicolumn{2}{c}{\VSgr} & \multicolumn{2}{c}{\VAnd} \\
                    & Observed & Predicted  & Observed & Predicted \\
\hline
$P_{\rm orb}$ (hr)  & $20.82$      & $20.85$   & $14.2582$  & $14.2639$   \\
R$_2$ (\Rsun)         & $<1.09$      & $1.076$   & $\sim0.8-1.0$  & $0.848$ \\
M$_2$ (\Msun)         & $<0.24$      & $0.233$   & $\sim0.2-0.4$  & $0.242$ \\
T$_{\rm eff,2}$ (K)   & $4900-5250$  & $5138$    & $5100\pm200$  & $5006$ \\
\hline
\end{tabular}
\end{table}


The evolution of the post-CE binaries leading to the present-day configurations of both \VSgr~and \VAnd~is rather similar.
In particular, we can explain both systems with an initial post-CE binary having a WD mass of $0.8$~\Msun~and a main-sequence star companion of mass $1.1$~\Msun.
However, as the observed orbital period of \VSgr~is longer than that of \VAnd, the orbital period of the initial post-CE binary has to be longer for the former ($3.5$~d) in comparison to the latter ($3$~d).
Therefore, the slight differences observed in both systems can be explained as the evolutionary stage of the companion of the WD at the onset of mass transfer.
Right after CE evolution, the orbital period is too long for the companion of the WD to fill its Roche lobe as a main-sequence star.
Therefore, it has to first become a subgiant before eventually filling its Roche lobe due to the expansion driven by hydrogen shell burning, as illustrated in Figs.~\ref{Fig:LvsT} and \ref{Fig:RvsT}.
This detached post-CE evolution takes several Gyr (Tables~\ref{Tab:FormationChannelV1082Sgr} and \ref{Tab:FormationChannelV479And}).

After the onset of mass transfer, although the donor star can maintain its hydrostatic equilibrium, it must sacrifice its thermal equilibrium in response to Roche-lobe-filling mass transfer.
This causes the mass-loss timescale to be comparable to the thermal (Kelvin–Helmholtz) timescale of the donor star.
During thermal timescale mass transfer, the mass transfer rates are so high (${\gtrsim10^{-7}}$~\Msunyr) that hydrogen can be burnt steadily on the surface of the WD, causing an increase in its mass, as illustrated in Figs.~\ref{Fig:PvsM} and \ref{Fig:PvsMdot}.
As the mass of the donor star decreases, at some point the mass-loss timescale starts to increase and becomes greater than the thermal timescale of the donor star.
When this happens, the donor can reestablish thermal equilibrium, and the thermal timescale mass transfer ends.

From this point on, the mass transfer rate decreases as the donor mass decreases since the evolution becomes driven by magnetic braking, and hydrogen burning on the surface of the WD is unstable, leading to cyclic nova eruptions.
The accreted matter is then ejected from the binary due to these repetitive nova eruptions, which keep the WD mass roughly constant.
Meanwhile, the binary evolves towards shorter periods as the donor mass decreases.
At present, the binary has properties comparable to those observed (Table~\ref{Tab:Comparison}).

\section{Discussion}

A new data set was collected on the two extremely long-period CVs V1082\,Sgr and V479\, including TESS-photometry, UV and IR spectroscopy. Combined with the circular polarimetry recently reported by \citet{2025ApJ...984..152L}, allows us to better interpret the controversial nature of these systems and carry out detailed evolutionary models to account for their characteristics. 

The \textit{TESS} data helped to reveal periodic variability in the light curves of V479\,And. The double orbital frequency, prominent in only one block of TESS data, is evidence of a double-hump light curve per orbital period, characteristic of ellipsoidal distortion of the Roche-lobe-filling donor star. 
The donor star is a significant contributor to light at the near-IR wavelength, corresponding to the TESS bandpass.  However, a small amplitude of variability $ \leq1.5$\% results either from a star not completely filling its Roche Lobe, or a very small angle of the binary orbit. We show in Section\,\ref{sec:inc} that the inclination angle is indeed small from other considerations. 

The UV spectrum of V479\,And revealed an anomalously high \ion{N}{v}/\ion{C}{iv} emission-line flux ratios. This is a common feature among CVs with evolved donors \citep{Gan2003}. A more extensive selection of such systems with nuclearly evolved donors is presented by \citet{2015ApJ...815..131K}. V479\,And with \ion{N}{v}/\ion{C}{iv} = 2.8 falls within a cluster of CVs of inverted line ratio systems BY\,Cam, EY\,Cyg, BZ\,UMa, EI\,Psc, and AE\,Aqr \citep[][their fig.~12]{2015ApJ...815..131K}. It confirms that V479\,And had an alternative evolutionary path in comparison to typical CV evolution with unevolved donor stars. 

Meanwhile, the circular polarization data of V1082\,Sgr revealed the presence of significant power at f$=12.352$\,d$^{-1}$ (i.e., 1.943\,hr, \citealt{2025ApJ...984..152L}). By detecting the circular polarization modulated with an $\approx0.1\times$ P$_{\mathrm {orb}}$, it was firmly established that V1082\,Sgr is indeed an intermediate polar with the third longest known spin period.
This, alone, essentially invalidates the previously proposed mass transfer model governed by the coupling magnetic fields \citep{Tovmassian_2016}. That model already had difficulties explaining a high mass transfer rate from the stellar wind. This system is better explained by assuming it is semi-detached, with mass transfer occurring through the \( L_1 \) point from the Roche-filling secondary. 

In addition, the IR spectroscopy showed an anomalous chemical composition of the donor star in V1082\,Sgr, identical to some other CVs \citep{2016ApJ...833...14H}. Much like the similarly divergent chemical composition of accretion material of V479\,And it is strong proof that V1082\,Sgr went through a TTMT evolutionary episode, as proposed by \citet{2019MNRAS.489.3031X}.

We revisited the basic binary parameters of both systems. We showed in Sect.~\ref{sec:R2M2} that both V479\,And and V1082\,Sgr contain radically lightweight donor stars. This confirms and strengthens the argument that their secondary stars are not only nuclearly evolved but also have lost most of their envelopes. It makes their further evolution even more intriguing.

The scenario according to which the companion of V1082\,Sgr has evolved off the main sequence and lost ${\sim0.5-2.0}$\,\Msun\ of its mass was initially proposed by \citet{2019MNRAS.489.3031X} for V1082\, Sgr. As a result of their numerical calculations, the companion appears as a K-type star with an effective temperature of 4660–5100\,K and mass ranging from M$_2=0.45 - 0.568$\,\Msun. However, this is inconsistent with the parameters of the donor star we derived here (Fig.\,\ref{fig:r2m2}).
We calculated evolutionary models, achieving a satisfactory agreement between the predicted and observed properties of the donor stars, specifically their masses, radii, and luminosities.

After CE evolution, the evolution of the binary can be driven by either nuclear evolution, or angular momentum loss due to magnetic braking, or a combination of both.
While the nuclear timescale evolution is relatively well understood, the same cannot be said about magnetic braking.
The strength and main dependencies of magnetic braking on stellar parameters (e.g. mass, structure, magnetic field, and evolutionary stage) are highly uncertain.
Despite that, there has been progress in understanding magnetic braking in the past years.

Regarding unevolved main-sequence stars, there is growing evidence that magnetic braking saturates \citep[e.g.,][]{Reiners_2009}, that is, beyond a certain rotation rate, the activity becomes independent of the rotation rate.
This means that the angular momentum loss due to magnetic braking becomes only weakly dependent on the rotation rate \citep[e.g.,][]{Chaboyer_1995}.
This also occurs for stars in binary main-sequence stars as well as detached and semi-detached post-CE WD binaries \citep[e.g.,][]{ElBadry_2022,Belloni_2024,BarrazaJorquera_2025}.
Additionally, magnetic braking is much more efficient for stars with a radiative core and a convective envelope in comparison to fully convective stars \citep{Belloni_2024,Schreiberetal24-1,BarrazaJorquera_2025}.

On the other hand, predictions for evolved main-sequence stars and subgiants are in better agreement with observations when magnetic braking is assumed to be much stronger in comparison to unevolved main-sequence stars.
For instance, the characteristics of the populations of transient and persistent neutron star low-mass X-ray binaries, transitional millisecond pulsar binaries, and ultra-compact X-ray binaries are much better explained by assuming the CARB \citep{CARB} model (or other similar prescriptions) for strong magnetic braking \citep[e.g.][]{Van_2019,CARB,Van_2021,Deng_2021,Chen_2021,Shahbaz_2022,Wei_2023,CastroSegura_2024,Echeveste_2024,Cui_2024,Yang_2024,Kar_2025,Yang_2025}.
Similarly, the fine-tuning problem related to the formation of ultra-compact X-ray binaries, detached close millisecond pulsar binaries, and AM\,CVn binaries can be solved by assuming strong magnetic braking \citep[e.g.][]{Chen_2021,Soethe_2021,BelloniSchreiber_2023}.
Here, we provide further evidence for strong magnetic braking acting on subgiants, which is required to explain the orbital periods of both \VAnd~and \VSgr, whose donor stars have undergone CNO processing.

We performed new evolutionary calculations adopting the CARB model \citep{CARB} for magnetic braking.
Due to the high AML rates predicted by this model, the binary evolution after CE evolution is convergent, despite the donor star having already evolved into a subgiant.
The introduction of the CARB in our modeling is the noteworthy difference between our models and those calculated by \citet{2019MNRAS.489.3031X}.

Despite the success of the CARB model when applied to nuclear evolved donors in accreting compact objects, it also faces problems.
For example, \citet{Fan_2024} showed that the CARB model cannot explain the orbital-period derivative of two neutron star low-mass X-ray binaries (SAX~J1748.9-2021 and XTE~J1710-281) and two black hole low-mass X-ray binaries (A0620-00 and  Nova Muscae 1991).
More recently, \citet{Belloni_2025} showed that the paradoxical system \paradoxical, which is a double WD binary with an orbital period of only 4.56~hr, can only be explained if the post-CE evolution is convergent, requiring an AML rate that is ${\sim100}$ times higher than the predicted by the CARB model.
That being said, although the CARB model is a promising model for nuclear evolved stars, it still needs improvements to account for several difficulties.

\section{Conclusions}

We have collected a wealth of new data on two unusual CVs, V1082\,Sgr and V479\,And,  with extremely long orbital periods, 20.8 and 14.3 hours, respectively, compared to most CVs.
We have used these data to determine the properties of the secondary stars in the two systems. As detailed in Table\,\ref{Tab:Comparison}, both have secondary star masses of about 0.24\,\Msun, and effective temperatures of about 5100\,K. V479\,And has a smaller donor star radius of about 0.85\,\Rsun, while V1082\,Sgr is about 1.1\,\Rsun, as a result of its longer period.
Curiously, we find that both objects have very low inclination.  This explains the low amplitude of the double-hump variability of the light curve of V479\,And ({\textit {TESS}}) data described here, and non-detection in V1082\,Sgr \citep{Tovmassian_2018b}.

Recent study has established that V1082\,Sgr hosts a moderate strength magnetic white dwarf and is classified as an intermediate polar \citep{2025ApJ...984..152L}. 
For V479\,And, we report the presence of an intense \ion{He}{ii} line in its UV spectrum, which is generally regarded as indirect evidence of a strongly magnetic white dwarf, consistent with the conclusion of \citet{2013A&A...553A..28G} who identified repetitive X-ray flashes as a signature of a polar. 
Our analysis shows that, despite these differences in magnetic properties and classification, both binaries appear to follow a similar evolutionary trajectory. 
This suggests that the secular evolution of these long-period cataclysmic variables is not strongly dependent on whether the accreting white dwarf is an intermediate polar or a polar candidate.

Furthermore, we found that the donor stars in both systems have atypical chemical compositions and significantly lower masses than most objects in their spectral classes.
These differences in characteristics compared to most CVs lead us to conclude that a brief and unstable phase of mass transfer has been involved in the evolution of both systems.
Our evolutionary models explain the diverging chemical composition of the emitting gas in these systems compared to the normal CVs.
By simulating binary evolution models that assume strong magnetic braking, we achieved a better agreement with observations, considering the measured distances, luminosities, and radii of donor stars determined independently from observations, in comparison to previous models.

Our findings provide further evidence for stronger magnetic braking acting on subgiants in close-orbit accreting compact objects in comparison to unevolved main-sequence stars.
Among accreting white dwarfs, they might significantly contribute to the population of close binary white dwarfs.
To further test this hypothesis,  similarly detailed investigations of other extremely long-period systems are required.

\begin{acknowledgements}
We thank the referee of this paper, Dr Albert Bruch, for his detailed reading and insightful comments, which enhanced the clarity and rigor of the manuscript. GT was supported by grants IN109723 from the Programa de Apoyo a Proyectos de Investigación e Innovación Tecnológica (PAPIIT).
DB acknowledges support from the São Paulo Research Foundation (FAPESP), Brazil, Process Numbers {\#2024/03736-2} and {\#2025/00817-4}. This project has received funding from the European Research Council (ERC) under the European Union’s Horizon 2020 research and innovation programme (Grant agreement No. 101020057). This research was supported by Deutsche Forschungsgemeinschaft  (DFG, German Research Foundation) under Germany’s Excellence Strategy - EXC 2121 "Quantum Universe" – 390833306. Co-funded by the European Union (ERC, CompactBINARIES, 101078773) Views and opinions expressed are, however, those of the author(s) only and do not necessarily reflect those of the European Union or the European Research Council. Neither the European Union nor the granting authority can be held responsible for them. Partial support for KSL's effort on the project was provided by NASA through grant numbers HST-GO-16489 and HST-GO-16659 from the Space Telescope Science Institute, which is operated by AURA, Inc., under NASA contract NAS 5-26555. RIH acknowledges support from NASA through grant number HST-GO-15454 from the Space Telescope Science Institute, which is operated by AURA, Inc., under NASA contract NAS 5-26555. 

\end{acknowledgements}

\bibliographystyle{aa} 
\bibliography{references} 

%
%
\begin{appendix}

\section{Secular evolution modeling of V479\,And and V1082\,Sgr}
\label{app:ModelingMESA}

\begin{table}
\caption{Adopted stellar and binary evolution parameters in MESA.}
\label{Tab:Assumptions}
\centering
\setlength\tabcolsep{0.05pt} 
\renewcommand{\arraystretch}{1.75} 
\begin{tabular}{lc}
\hline
Parameter & Value \\
\hline
initial post-CE orbital period &  $2.5,\,\textbf{\color{Green} 3.0},\,\textbf{\color{magenta} 3.5}$~d \\
initial post-CE white dwarf mass &  $0.6,\,0.7,\,\textbf{0.8},\,0.9,\,1.0$~\Msun \\
initial post-CE companion mass &  $1.0,\,\textbf{1.1},\,1.2,\,1.3,\,1.4,\,1.5$~\Msun \\
Metallicity Z  &  $\textbf{\color{magenta} 0.010},\,\textbf{\color{Green} 0.015},\,0.020,\,0.025$ \\
criterion for stable H shell burning & \citet{Wolf_2013} \\
MB prescription  &  CARB \citep{CARB} \\
CAML prescription  &  mass loss \citep{King_1995} \\
Reimers's parameter &  0.5 \\ 
MLT & \citet{Henyey_1965} \\
mixing length ($H_p$) & $\textbf{1.5},\,2.0,\,2.5$ \\
\vspace{-0.2cm}
extent of diffusive & \multirow{2}{*}{$0.016\,H_p$}\\
exponential core overshooting &  \\
\vspace{-0.2cm}
criterion for stability & \multirow{2}{*}{Schwarzschild $(\nabla_{\rm ad} = \nabla_{\rm rad})$} \\
against convection & \\
\vspace{-0.2cm}
\multirow{2}{*}{opacity} &  \textbf{\citet{Ferguson2005}}, \\
& \citet{Freedman2008,Freedman2014} \\
nuclear network & \texttt{cno$\_$extras.net} \\
atmosphere boundary conditions & Eddington T(tau) relation \\
\vspace{-0.2cm}
how opacities are calculated  & \multirow{2}{*}{\textbf{varying}, iterated} \\
throughout the atmosphere & \\
\hline
\end{tabular}
\begin{tablenotes}
\item[] \textbf{Notes.} When more than one value of the parameter is varied in the search, the values adopted for the best-fitting models (i.e. Table~\ref{Tab:Comparison}) are highlighted in boldface (magenta for \VSgr, green for \VAnd, and black for both).
%
%
\end{tablenotes}
\end{table}

We used the version r15140 of the MESA code \citep[][]{Paxton2011,Paxton2013,Paxton2015,Paxton2018,Paxton2019,Jermyn2023} to calculate post-common-envelope (CE) binary evolution from the moment just after CE evolution onward.
We describe here our assumptions for single star and binary evolution, which are summarized in Table~\ref{Tab:Assumptions}.
For reference, our approach follows closely that by \citet{BelloniSchreiber_2023}\,\footnote{\href{https://zenodo.org/records/8279474}{https://zenodo.org/records/8279474}}.

The MESA equation of state is a blend of the OPAL \citep{Rogers2002}, SCVH \citep{Saumon1995}, FreeEOS \citep{Irwin2004}, HELM \citep{Timmes2000}, PC \citep{Potekhin2010} and Skye \citep{Jermyn2021} equations of state.
Nuclear reaction rates are a combination of rates from NACRE \citep{Angulo1999}, JINA REACLIB \citep{Cyburt2010}, plus additional tabulated weak reaction rates \citep{Fuller1985,Oda1994,Langanke2000}.
Screening is included via the prescription of \citet{Chugunov2007} and thermal neutrino loss rates are from \citet{Itoh1996}.
Electron conduction opacities are from \citet{Cassisi2007} and radiative opacities are primarily from OPAL \citep{Iglesias1993,Iglesias1996}, with high-temperature Compton-scattering dominated regime calculated using the equations of \citet{Buchler1976}.

%
We varied the metallicity by adopting the values from ${Z=0.005}$ to $0.025$, in step of $0.005$, to account for sub-solar, solar, and super-solar metallicity models.
We assumed the grey Eddington T(tau) relation to calculate the outer boundary conditions of the atmosphere \citep[][their sect.~5.3]{Paxton2011}.
We adopted either a uniform opacity that is iterated to be consistent with the final surface temperature and pressure at the base of the atmosphere, or a varying opacity consistent with the local temperature and pressure throughout the atmosphere.
For low-temperature radiative opacities, we adopted those from either \citet{Ferguson2005} or \citet[][]{Freedman2008,Freedman2014}.
We further used the nuclear network \texttt{cno$\_$extras.net}, which accounts for the nuclear reactions of the carbon-nitrogen-oxygen hydrogen burning cycle.

%
We allowed the stars to lose mass through winds, adopting the \citet{Reimers_1975} prescription, setting the wind efficiency to $0.5$.
%
For the evolutionary phases with convective core, that is, core hydrogen and helium burning, we took into account exponential diffusive overshooting, assuming a smooth transition in the range ${1.2-2.0}$~\Msun~\citep[e.g.,][]{Anders_2023}.
We assumed that the extent of the overshoot region corresponds to ${0.016~H_{\rm p}}$ \citep[e.g.,][]{Schaller_1992,Freytag_1996,Herwig_2000}, with $H_{\rm p}$ being the pressure scale height at the convective boundary.
We treated convective regions using the scheme by \citet{Henyey_1965} for the mixing-length theory, varying the mixing length (in units of $H_{\rm p}$): $1.5,2.0,2.5$ \citep[e.g.,][]{Joyce_2023}.

%
The Roche-lobe radius of each star was computed using the fit of \citet{eggleton1983}.
The mass transfer rates due to Roche-lobe overflow are determined following the prescription of \citet{Ritter1988}, in which the atmosphere of the star is filling the Roche lobe.
In this so-called atmospheric Roche-lobe overflow model, mass transfer occurs even when the star radius is smaller than the Roche-lobe radius.

%
We enforced that the WD accretor can only accrete a part of the mass being transferred from its companion.
Depending on the accretion rate onto the WD, hydrogen shell burning could be stable, which increases its mass.
We implemented the critical accretion rate $\dot{M}_{\rm stable}$ calculated by \citet[][]{Wolf_2013}, above which WDs are thermally stable, that is, hydrogen burns steadily in a shell.
For accretion rates lower than this critical value, the WD undergoes nova eruptions, such that all of the accreted mass is assumed to be expelled from the binary \citep[although it is unlikely that exactly 100\% of the mass is ejected, e.g.][]{Jose_2020}.
We further assumed that there is a maximum possible accretion rate $\dot{M}_{\rm red-giant}$ \citep[][]{Wolf_2013} such that WDs accreting at rates above it will burn stably at this maximum rate, and the remaining non-accreted matter will be piled up forming a red-giant-like envelope, which is assumed to be lost from the binary in the form of stellar-like winds.

Regarding orbital angular momentum loss, we assumed for the magnetic braking the CARB  prescription \citep{CARB} given by

\begin{equation}
\begin{multlined}
\dot{J}_{\rm MB} \ = \ 
- \, 2\times10^{-6}
\left(
   \frac{-\dot{M}_{\rm wind}}{\rm g~s^{-1}}
\right)^{-1/3}
\left(
   \frac{R}{\rm cm}
\right)^{14/3}
\left(
   \frac{\Omega}{\Omega_\odot}
\right)^{11/3} \, \times
\\
\left(
    \frac{\tau_{\rm conv} }{\tau_{\odot, \rm conv}}
\right)^{8/3}
\left[
  \left(
      \frac{v_{\rm esc}}{\rm cm~s^{-1}}
  \right)^2\,+\,\frac{2}{K_2^2}\,
  \left(
      \frac{\Omega}{\rm s^{-1}}
  \right)^2
  \left(
      \frac{R}{\rm cm}
  \right)^2
\right]^{-2/3}
\end{multlined}
\label{Eq:MB-CARB}
\end{equation}

\noindent
where $\dot{M}_{\rm wind}$, $R$, $\Omega$, $\tau_{\rm conv}$ are the wind mass-loss rate, radius, spin and convective turnover timescale of the companion of the white dwarf, respectively.
The convective turnover timescale was calculated by integrating the inverse of the velocity of convective cells, as given by the mixing-length theory, over the radial extent of the convective envelope.
The Sun spin and convective turnover timescale are  ${3\times10^{-6}}$~${\rm s^{-1}}$ and ${2.8\times10^6}$~s, respectively, and ${K_2=0.07}$.
Finally, $v_{\rm esc}$ is the escape velocity.

We started all our simulations immediately after the white dwarf formation, that is, we assume as initial configuration a detached post-CE binary consisting of a point-mass white dwarf and a zero-age main-sequence star. 
The initial parameters of the post-CE binaries covered white dwarf masses from $0.6$ to $1.0$~\Msun, companion masses from $1.0$ to $1.5$~\Msun, and orbital periods ranging from ${\sim0.1}$ to ${\sim5}$~d.
We would like to emphasize that starting with zero-age main-sequence stars does not compromise our results.
This is because, given their initial masses, they will spend at least a few Gyr on the main sequence, while the typical timescale for the formation of an average cataclysmic variable white dwarf are much shorter \citep[${\sim300}$~Myr,][]{Belloni_2018b}.

\section{Formation Pathways for of V479~And and V1082~Sgr}
\label{app:FormationChannels}

\begin{table*}
\centering
\caption{Evolution of a zero-age post-CE binary towards the present-day properties of \VSgr.}
\label{Tab:FormationChannelV1082Sgr}
\setlength\tabcolsep{9pt} 
\renewcommand{\arraystretch}{1.25} 
\begin{tabular}{r c c c c c l}
\hline
\noalign{\smallskip}
 Time  &   $M_1$    &   $M_2$    & Type$_1$  & Type$_2$ & Orbital Period & Event\\
 (Myr) & (M$_\odot$)&(M$_\odot$) &           &          &  (d)        &      \\
\hline
\noalign{\smallskip}
     0.0000  &  0.800  &  1.100  & WD     & MS  &    84.0000  &  initial post-common-envelope binary \\
  5011.2289  &  0.800  &  1.098  & WD     & SG  &    84.1151  &  secondary becomes a subgiant \\
  6492.0269  &  0.800  &  1.096  & WD     & SG  &    28.2261  &  begin RLOF\\
\textbf{6516.0473}  &  \textbf{1.010}  &  \textbf{0.233} & WD     & \textbf{SG}       &    \textbf{20.8487}  &  \textbf{Binary looks like \VSgr} \\
6864.6246  &  1.010  &  0.176 & WD     & proto-WD &    19.0684  &  end RLOF \\
\noalign{\smallskip}
\hline
\end{tabular}
\begin{tablenotes}
\item[] \textbf{Notes.} The quantities $M_1$ and $M_2$ and Type$_1$ and Type$_2$ are the masses and stellar types of the progenitors of the WD and its companion, respectively. $P_{\rm orb}$ is the orbital period and the last column corresponds to the event occurring to the binary at the given time in the first column.
The row in which the binary has the present-day properties of \VSgr~is highlighted in boldface.
Abbreviations:
MS~(main~sequence~star),
SG~(subgiant~star),
WD~(white~dwarf),
RLOF~(Roche~lobe~overflow).
\end{tablenotes}
\end{table*}

\begin{table*}
\centering
\caption{Evolution of a zero-age post-CE binary towards the present-day properties of \VAnd.}
\label{Tab:FormationChannelV479And}
\setlength\tabcolsep{9pt} 
\renewcommand{\arraystretch}{1.25} 
\begin{tabular}{r c c c c c l}
\hline
\noalign{\smallskip}
 Time  &   $M_1$    &   $M_2$    & Type$_1$  & Type$_2$ & Orbital Period & Event\\
 (Myr) & (M$_\odot$)&(M$_\odot$) &           &          &  (h)        &      \\
\hline
\noalign{\smallskip}
   0.0000  &  0.800  &  1.100 & WD     & MS       &    72.0000  &  initial post-common-envelope binary \\
5457.1641  &  0.800  &  1.098 & WD     & SG       &    72.0380  &  secondary becomes a subgiant \\
7129.2901  &  0.800  &  1.096 & WD     & SG       &    25.1126  &  begin RLOF\\
\textbf{7147.7745}  &  \textbf{0.942}  &  \textbf{0.242} & \textbf{WD}     & \textbf{SG}       &    \textbf{14.2639}  &  \textbf{Binary looks like \VAnd} \\
7573.8580  &  0.942  &  0.165 & WD     & proto-WD &    11.9338  &  end RLOF \\
\noalign{\smallskip}
\hline
\end{tabular}
\begin{tablenotes}
\item[] \textbf{Notes.} The quantities $M_1$ and $M_2$ and Type$_1$ and Type$_2$ are the masses and stellar types of the progenitors of the WD and its companion, respectively. $P_{\rm orb}$ is the orbital period and the last column corresponds to the event occurring to the binary at the given time in the first column.
The row in which the binary has the present-day properties of \VAnd~is highlighted in boldface.
Abbreviations:
MS~(main~sequence~star),
SG~(subgiant~star),
WD~(white~dwarf),
RLOF~(Roche~lobe~overflow).
\end{tablenotes}
\end{table*}


The formation pathways we obtained for \VSgr~and \VAnd~are given in Tables~\ref{Tab:FormationChannelV1082Sgr} and \ref{Tab:FormationChannelV479And}, respectively.

\end{appendix}

\end{document}